\documentclass[graybox, secnum]{svmult}

\usepackage{mathptmx}       
\usepackage{siunitx}
\usepackage{newtxtext,newtxmath}
\usepackage{helvet}         
\usepackage{courier}        
\usepackage{type1cm}        
\usepackage{makeidx}         
\usepackage{graphicx}        
\usepackage{multicol}        
\usepackage[bottom]{footmisc}
\usepackage[unicode]{hyperref}        
\usepackage{soul}            
\hypersetup{colorlinks=true,urlcolor=blue}
\usepackage[square,numbers]{natbib}
\makeindex             

\begin{document}
\title*{Lobster Eye X-ray Optics}
\author{Rene Hudec \thanks{corresponding author} and Charly Feldman}
\institute{Rene Hudec \at Czech Technical Universiuty in Prague, Technicka 2, CZ-160 00 Praha 2, Czech Republic \email{hudecren@fel.cvut.cz}
\and Charly Feldman \at School of Physics and Astronomy
University of Leicester, University Road, LE1 7RH, UK \email{chf7@leicester.ac.uk}}
%
%
\maketitle
\abstract{This chapter describes the history, principles, and recent developments of large field of view X-ray optics based on lobster eye designs.
Most of grazing incidence (reflective) X-ray imaging systems used in astronomy and other applications, are based on the Wolter 1 (or modified) arrangement. But there are also other designs and configurations proposed for future applications for both laboratory and space environments. Kirkpatrick-Baez (K-B) based lenses as well as various types of lobster eye optics serve as an example. Analogously to Wolter lenses, all these systems use the principle that the X-rays are reflected twice to create focal images.
Various future projects in X-ray astronomy and astrophysics will require large optics with wide fields of view. Both large Kirkpatrick-Baez modules and lobster eye X-ray telescopes may serve as solutions as these can offer innovations such as wide fields of view, low mass and reduced costs. The basic workings of lobster eye optics using Micro Pore Optics (MPOs) and their various uses are discussed. The issues and limiting factors of these optics are evaluated and current missions using lobster eye optics to fulfill their science objectives are reviewed. The Multi Foil Optics (MFO) approach represents a promising alternative. These arrangements can also be widely applied in laboratory devices.
The chapter also examines the details of alternative applications for non-Wolter systems in other areas of science, where some of these systems have already demonstrated their advantages such as the K-B systems which have already found wide applications in laboratories and synchrotrons.}

\section{Keywords} 
X--ray astronomy, X--ray astrophysics, X--ray optics, X-ray telescope, lobster eye optic, Micro Pore Optics

\section{Introduction}
In this chapter, we focus on the non-Wolter grazing incidence X-ray imaging systems with emphasis on lobster eye (LE) wide Field of View (FoV) systems.

Many scientific achievements over the last two decades in X-ray astronomy are closely related to the use of imaging X-ray telescopes. In astronomy and astrophysics, it was the use of imaging X-ray telescopes based on grazing incidence X-ray optics that opened a completely new window into the Universe and has lead to great discoveries during past decades. To acknowledge these achievements, the 2002 Nobel Prize for physics was awarded to Professor Riccardo Giacconi who significantly contributed to the construction of the first astronomical X-ray telescopes in the 60's and 70's. These telescopes achieve much better signal to noise ratio than X-ray experiments without optics, which allows for the detection of faint sources for example. The use of X-ray optics further allows imaging, precise localization, photometry, spectroscopy, variability studies, and an estimation of physical parameters of X-ray emitting regions (temperature, electron density...). Space borne X-ray optics are also well suited for monitoring the X-ray sky for variable and transient objects including X-ray novae, X-ray transients, X-ray flares of stars and AGNs, galactic bulge sources, X-ray binaries, SGRs (Soft Gamma Ray Repeaters) and X-ray afterglows of GRBs (Gamma Ray Bursts). X-ray optics are an important part of numerous past, recent, and future space projects (EXOSAT, ROSAT, Einstein, RT-4M Salyut 7, Fobos, AXAF/Chandra, XMM-Newton, ABRIXAS, BeppoSAX, ASCA, XEUS, Athena...). 

In the laboratory, there are numerous applications for X-ray optics, e.g. in plasma physics, laser plasma, synchrotron analyses, biology, crystallography, medicine, material and structure testing, X-ray lithography, etc.

This chapter reviews and discusses X-ray imaging mirrors based on grazing incidence reflections which are alternatives to Wolter systems, with an emphasis on lobster eye optics. These systems are described elsewhere in the literature, but their space and astronomy applications are still marginally discussed. We review and discuss these systems, their past and recent ground-based applications, and their potential for future X-ray astronomy and laboratory applications.

The grazing incidence reflecting X-ray lenses discussed in this work typically reflect from the optical to soft X-rays of 2 to 10 keV, depending on the reflecting surface material and on the angle of incidence. Since there are science goals that require higher energies, recent efforts have focused on various improvements and additional surface layers, such as multi-layers, to meet this.

We give a brief introduction of the various types of the lobster eye X-ray optics in Section \ref{intro} where both Schmidt and Angel optics are discussed. In Section \ref{secmpo}, the lobster eye systems in the Angel arrangement, based on Micro Pore Optics (MPOs) are described. In Section \ref{secsch}, the wide-field systems of lobster eye optics in the Schmidt arrangement based on the Multi Foil Technology (MFO) are discussed in detail, with emphasis on prototypes already designed, developed and tested. Section \ref{KBop} addresses K-B X--ray optics and finally in Section \ref{sum}, we give a summary of the chapter including a comparison between the MPO and MFO technologies.

\section{Lobster eye X-ray optics}
\label{intro}
\subsection{Introduction}
\begin{figure}
	\centering
		\includegraphics[width=\columnwidth]{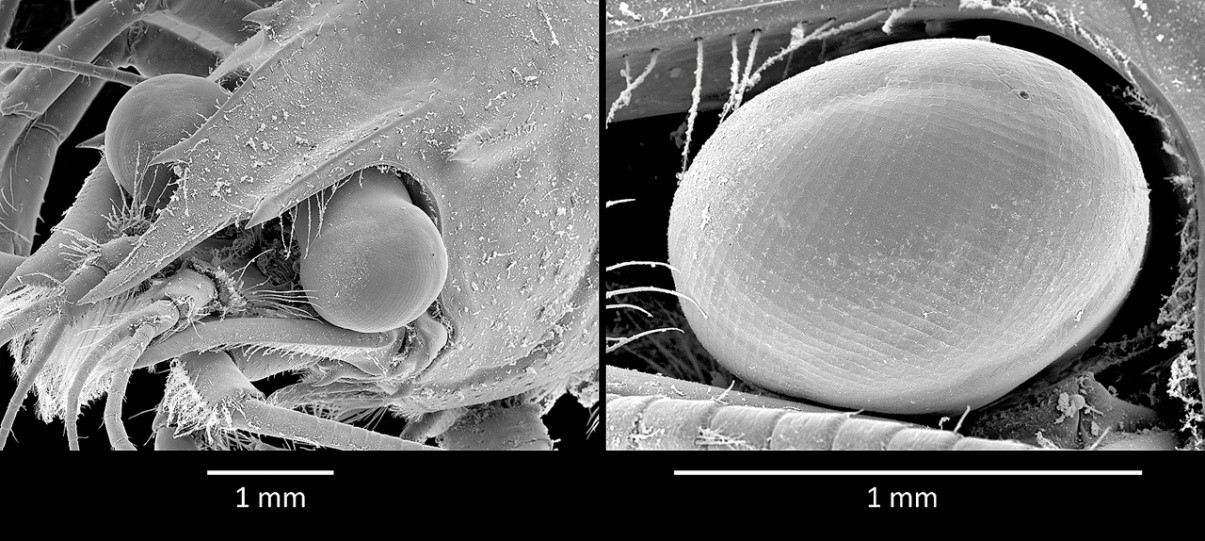}
		\includegraphics[width=0.58\textwidth]{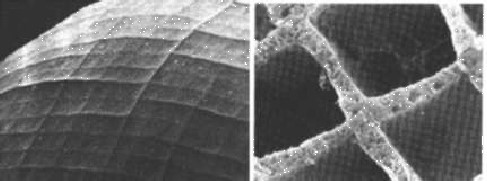}
		\includegraphics[width=0.40\textwidth]{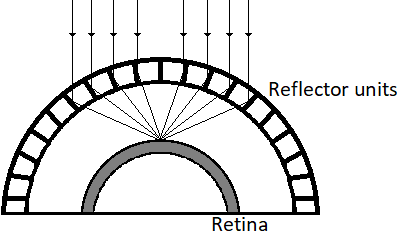}
			\caption{\textit{Above:} A lobster's eye - The head and a zoomed in image of the eye, and \textit{Below:} electron microscope images showing the surface details. A ray diagram showing the workings of the eye is shown bottom right.}
	\label{lob}
\end{figure}

Crustaceans eyes such as lobsters, shrimps and crayfish, provide an excellent opportunity for biomimicking and creating novel X-ray optics.
Instead of a standard lens, a lobster's eye consists of a large number of square pores evenly distributed across a sphere, with each pore pointed towards a common centre. The light reflects off the very smooth walls of each pore and is focused onto the curved retinal surface. The retina and the pores form concentric spheres with the radius of curvature of the retina being half that of the pores. The ratio of the width of the pores to their length requires the light to be reflected at very shallow angles - as is required to focus X-rays (see Figure \ref{lob}).

\begin{figure}
\begin{center}
\includegraphics[width=100mm]{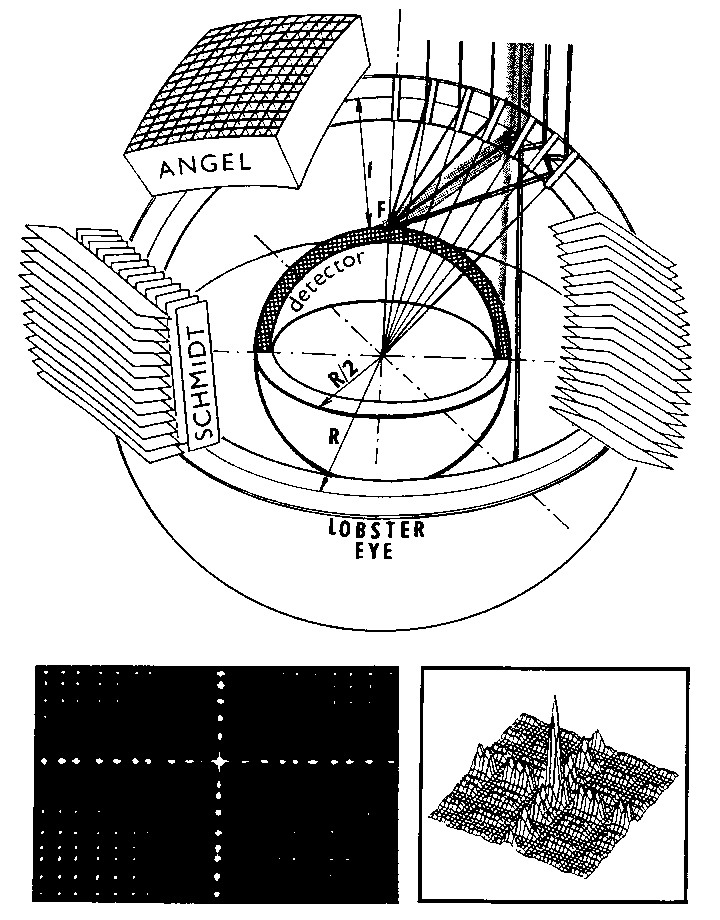}
\caption{\textit{Above:} The schematic arrangement of the lobster eye type X-ray optics in both Angel and Schmidt 2D and 1D geometries \textit{Below, left:} image produced by the Schmidt objective prototype in optical light and \textit{Below, right:} distribution of intensity on the focal sphere for a point-line source (computer ray-tracing) \cite{Inne01}.}
\label{figsch}
\end{center}
\end{figure}

The lobster eye geometry for X-ray imaging was first introduced by Angel in 1979\cite{ang}, where he discussed the possibility of creating an X-ray telescope with an extraordinarily large FoV, and even an all sky imager consisting of a full spherical optic and a spherical detector. The design is based on a modification of the K-B \cite{kirk} (see section \ref{KBop}) arrangement, but using successive orthogonal surfaces which are coincident in space to reflect the X-rays instead of separate successive surfaces. This dictates the need for pores with a square cross section - exactly mimicking a lobster's eye.
Angel's premise was that by replicating the geometry of crustaceans such as lobsters or crayfish, it would be possible to form an effective X-ray optic. The optics would need to be thin (1-2 mm thick), made up of regular, square cross section pores, which were evenly distributed over a spherical surface, with highly polished pore walls and an X-ray reflective coating. An alternative, but similar, arrangement had already been proposed by Schmidt in 1975\cite{Schmidt} (shown in Figure \ref{figsch}). Both of these arrangements will be discussed in full detail in the following sections.

The wide-field mirror modules offer advantageous application in astrophysics. The major scientific achievements of X-ray astronomy in the recent past are closely related to the use of large X-ray imaging telescopes based mostly on the Wolter 1 X-ray objectives. These systems usually achieve excellent angular resolution as well as very high sensitivity, but are quite limited in the FoV available, which is less than 1 degree in most cases.  However, the future of X-ray astronomy and astrophysics requires not only detailed observations of particular triggers, but also precise and highly sensitive X-ray sky surveys, patrol and monitoring. The  confirmed X-ray counterparts of GRBs may serve as an excellent example. Detailed investigations of GRBs, in almost all cases, indicate the presence of variable and/or fading X-ray counterparts/afterglows. The X-ray identification of GRBs has lead to great improvements in our study and understanding of these sources. This has enabled identifications at other wavelengths due to better localization accuracy provided in X-ray, compared with gamma ray observations. Since most GRBs seem to be accompanied by X-ray emission, the future systematic monitoring of these X-ray transients/afterglows is extremely important. However, these counterparts are faint in most cases, requiring powerful wide FoV telescopes. An obvious alternative seems to be the use of wide FoV X-ray optics allowing the signal to noise ratio to be increased if compared to non-focusing  optical systems. The desired limiting sensitivity for a lobster eye telescope is roughly  $10^{-12} erg^{-2}s^{-1}$ which would allow for the detection of an order of magnitude greater number of sources to be observed over the course of a day \cite{Sve04}. This is consistent with the fluxes associated with X-ray afterglows of GRBs.  Furthermore, the wide field X-ray telescopes play an important role in monitoring faint variable X-ray sources to provide better statistics of such objects (note e.g. the occurrence of two faint fading X-ray sources inside the gamma ray error box of GRB970616) as well as in other fields of X-ray astrophysics. The recent hunting for faint fading X-ray afterglows of GRBs has indicated that there is a large number of faint and/or variable X-ray sources worthy of detailed study.

There have been many attempts to increase the available FoV coverage of Wolter and analogous X-ray telescopes. To avoid any confusion, we suggest to restrict the term “wide-field X-ray optics” only to optical systems with a FoV significantly larger than 1 degree, whilst using the term “narrow-field system” for systems with FoV less than 1 degree.

In the following sections we describe in detail the two alternative approaches to the wide-field lobster eye optics, namely the Angel arrangement (based on MPO optics) and the Schmidt arrangement (based on MFO optics).

\section{Lobster eye telescopes using Micro Pore Optics}
\label{secmpo}

\subsection{Introduction}
As Angel\cite{ang} described, optics based on a lobster eye design could be used to create X-ray point-to-point imaging, focusing, collimation and beam splitters\cite{martin}. With a flat optic and equal optic-to-detector and optic-to-source distances, point-to-point imaging\cite{fraspie} of an unmagnified X-ray source is achievable. In this instance, the pores are parallel to one another and are arranged such that they are parallel to the line defined by the centre of the source to the centre of the detector. In this geometry, the expanding cone of X-rays from the source is focused onto the detector from a region of the optic defined by the critical angle for X-ray reflection. For imaging of sources, focusing is achieved when the optic is slumped with a spherical Radius of Curvature (RoC) and the detector is placed at the focal length where $F=RoC/2$. The convex surface of the optic is positioned towards the source\cite{fraspie2}. This geometry works for both finite and infinite source distances (e.g. planets, stars etc.) but where the optic to source distance is larger than the optic to detector distance. Collimation can be achieved by chemically roughening the channel walls for example, in this configuration a small FoV is defined, and rays from outside this filed of view are efficiently blocked, reducing the sky background\cite{Mineo}. Beam splitters are created by rotating a slumped optic so that the concave surface of the optic is towards the X-ray source. By setting the correct optic-to-source distance, a series of narrow parallel beams can be created, or by shortening the distance further, the beam can be radially and uniformly diverged. 

\begin{figure}
	\centering
		\includegraphics[width=0.31\textwidth]{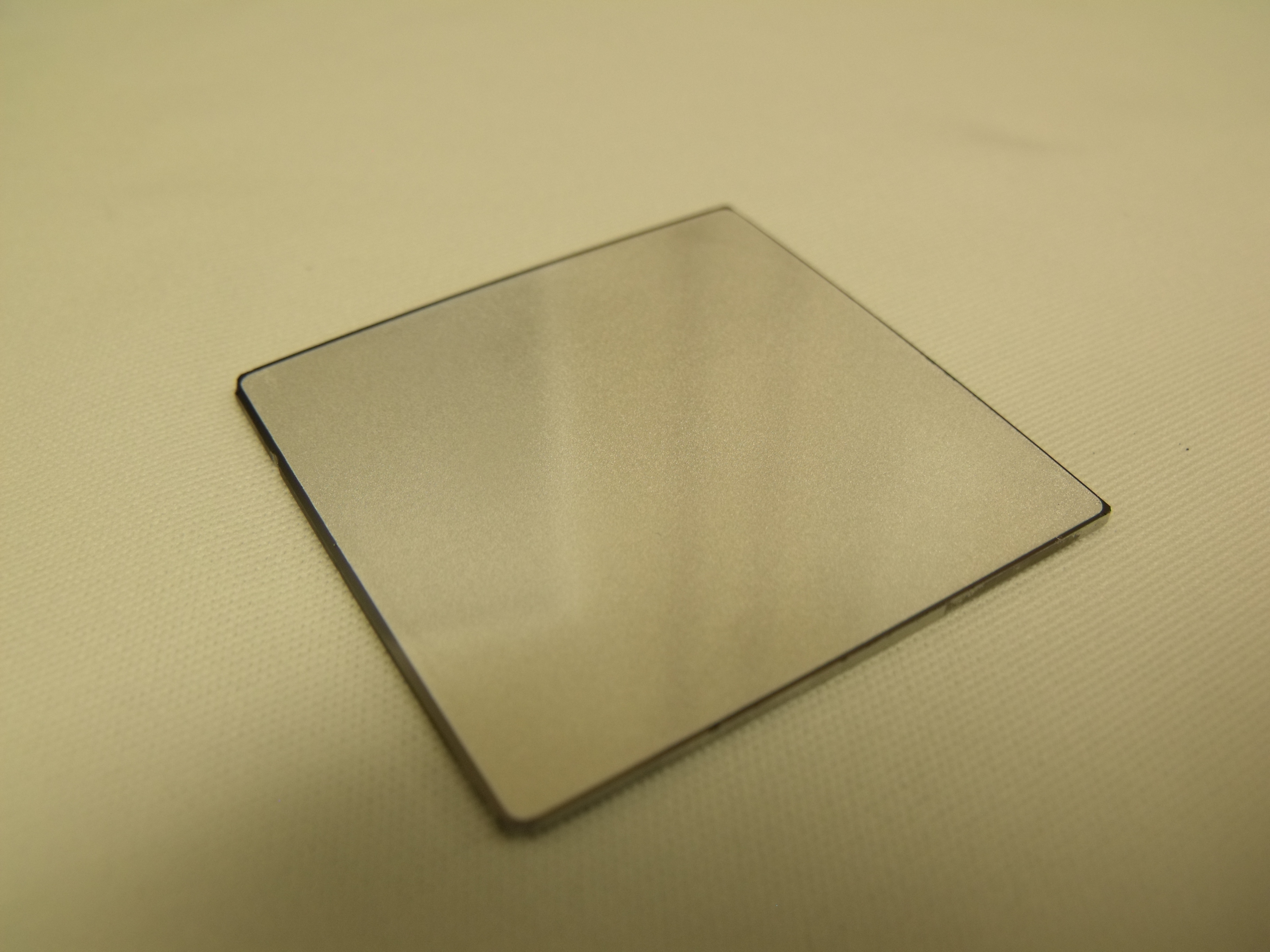}
		\includegraphics[width=0.31\textwidth]{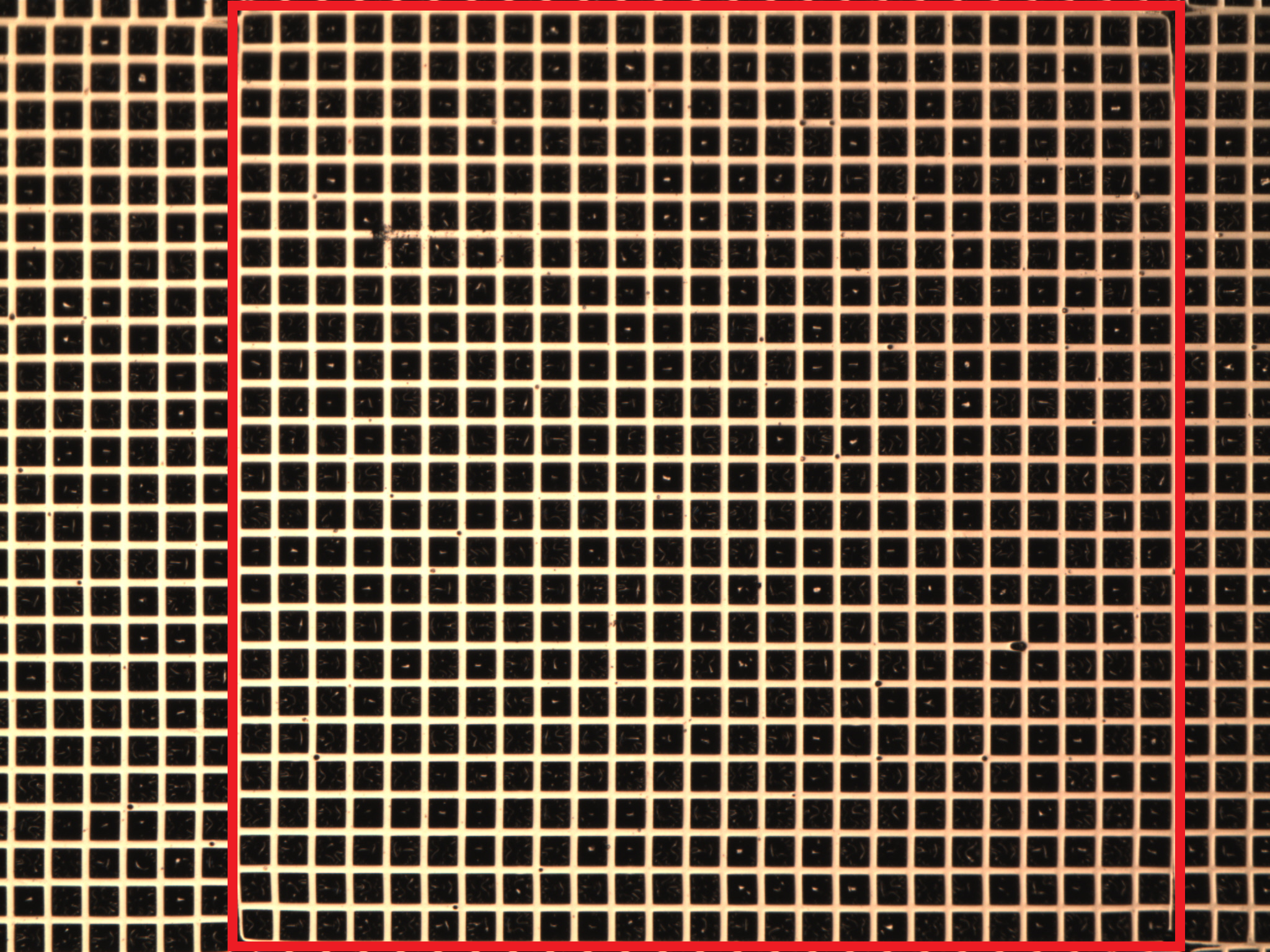}
			\caption{\textit{Left:} a single, 40 mm by 40 mm square, iridium coated, aluminium filmed, 1.05mm thick MPO Produced by Photonis France SAS. \textit{Right:} A microscope image of an MPO showing a single multifibre (enclosed within the red box) consisting of 25 by 25, 40 $\mu$m pores.}
	\label{figmpo}
\end{figure}

Micro Pore Optics (MPOs) are glass optics, typically 1-3 mm in thickness, with regular square pores. These pores are slumped to a spherical figure and coated to reflect X-rays. Usually, each MPO is 40 mm by 40 mm in size with 20 $\mu$m or 40 $\mu$m wide pores and with an open fraction of 60\% or higher. An example is shown on the left of Figure \ref{figmpo}.

Using MPOs in a lobster eye geometry, in comparison to a traditional Wolter telescope\cite{wolt}, can provide a very large FoV with very low mass. As the pore apertures are so small the channels do not have to be long to achieve the required grazing angles and therefore reduces the mass of the optic.

The size of the FoV of a lobster eye telescope depends only on the angular extent of the spherical optic and detector. A large lobster eye geometry can be realised by tessellating an array of smaller optic tiles, such as MPOs, over a larger spherical optic frame. If the frame has the same spherical figure as the slumped optics, then all channels will point to the centre of that sphere.
Providing the optic is constructed with very small gaps between the tessellated MPOs for support structures, there is little to no vignetting and no change in the PSF over the FoV of the assembled optic.

This optic design has been used and adapted for X-ray telescope missions such as the MXT on board SVOM\cite{gotz}, the SXI instrument on SMILE\cite{smile} and the WXT on the Chinese Einstein Probe\cite{ep} mission, and proposed missions including TAP\cite{tap} and Gamow\cite{gam}. The trade off with using this technology is the more modest angular resolution compared to traditional telescopes such as Chandra\cite{changman} and XMM-Newton\cite{2}. The Mercury Imaging X-ray Spectrometer (MIXS) instrument on BepiColombo\cite{mixs} used the same technology but in a different way. MIXS comprises of both a collimator and a telescope. The telescope uses optics with radially packed square pores, in concentric rings to approximate a Wolter geometry. Further details on MIXS are discussed in Section \ref{bep}.

Section \ref{secmpo} gives a description of the workings of MPOs; detailing production, the formation of the unique characteristic Point Spread Function (PSF) and the calculation of the fundamental parameters for the mission specific energy band. The details and methodology of how to formulate a narrow field lobster eye optic arrangement are explained. Some of the issues and limiting factors of these optics are evaluated and finally, the current missions using lobster eye optics to fulfil their science objectives are reviewed.

\subsection{MPOs - Production and design}
At the time of Angel's paper\cite{ang}, Micro Channel Plates (MCPs), thin square glass plates with regular and smooth circular pores, were in use as photo multipliers\cite{woodhead,ruggi}, electron detectors and X-ray detectors\cite{wiza,mcp}. The similarities between the proposed optics needed by Angel to form lobster eye optics and MCPs was recognised and has since been extensively pursued by several authors \cite{theory, wilks, fraspie, kaa}. In order to differentiate between the glass plates used for detectors and those used for optics, the term MPOs has been adopted for the latter, whilst the former retained the name MCPs.

\begin{figure}
	\centering
		\includegraphics[width=0.95\textwidth]{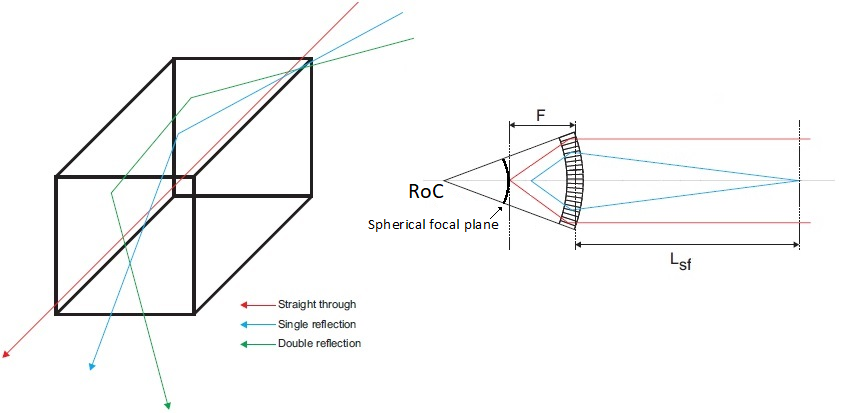}
			\caption{\textit{Left:} different ray paths through a single pore of an MPO. \textit{Right:} Geometry of a slumped MPO\cite{ady}. Red rays are from an infinite source and blue rays show the shortening due to a finite source distance. $L_{sf}$ is the finite source-to-optic distance, $F$ is the focal length, which is half of the RoC.}
	\label{geom1}
\end{figure}

The geometry of each MPO comprises a square packed array of microscopic pores, each with a square cross-section. The pores are arranged over a spherical surface with a RoC $R_{slump}$, where $F$ is the focal length of the MPO, as shown on the right of Figure \ref{geom1}. The ideal geometry is such that all the pores within the MPO point towards a common centre of curvature, with reflections from the walls of the pores producing an image on a spherical focal surface which is concentric with the spherical optic. The RoC of the spherical focal surface is equal to $F$, or half the RoC of the optic.

Whilst the technology of MPOs for X-ray optics (collimation or focusing) differs from MCP technology in several ways, they are both formed from lead glass and the manufacturing of the two is very similar. A full description of the production of MCPs and MPOs can be found in Feldman et al \cite{thspie}. MPOs consisting of square pores with widths of 10 $\mu$m up to 720 $\mu$m can be fabricated but for the majority of applications, they tend to be either 20 $\mu$m or 40 $\mu$m and are either square packed and slumped like a lobster eye, e.g. SVOM's MXT and SMILE's SXI, or radially packed, e.g. BepiColombo's MIXS-T.

The individual pore fibres are stacked to form square multifibres, the size of which are determined by the pore size. For example, an MPO made up of 40 $\mu$m pores will be stacked into square multifibres of 25 pores by 25 pores, which in turn will be stacked with $\sim$31 multifibres across in each direction to form an MPO of 40 mm by 40 mm. A single MPO formed as described with an aluminium film on the convex surface, is shown on the left of Figure \ref{figmpo}, and a microscope image of a complete multifibre is shown on the right.

The lead glass used to form the MPOs can effectively reflect incoming X-rays with a flat MPO providing a reflection efficiency of $\sim$60-80\% at Al-K (1.49 keV). However, a slumped bare glass MPO, which is required for X-ray focussing, only reflects $\sim$20-40\% (measured at Al-K, 1.49 keV, and an angle of 55 arcmin off-axis) of incomming X-rays. It is not fully understood why the reflection efficiency decreases during the slumping process. In order to improve the reflection of incident X-rays, the pores are coated with iridium (10 $\pm$ 5 nm thick for MIXS and 25 $\pm$ 5 nm for the MXT) by a process of atomic layer deposition. Studies are being carried out to investigate nickel, platinum and multilayer coatings, to improve the X-ray reflection efficiency at specific energy bands relevant to different mission's science goals. On the front (convex) surface of the MPO, an aluminium film of 60-100 nm thickness is applied. This layer acts as both an optical blocking filter and as a thermal control surface.

\begin{figure}
	\centering
		\includegraphics[width=0.75\textwidth]{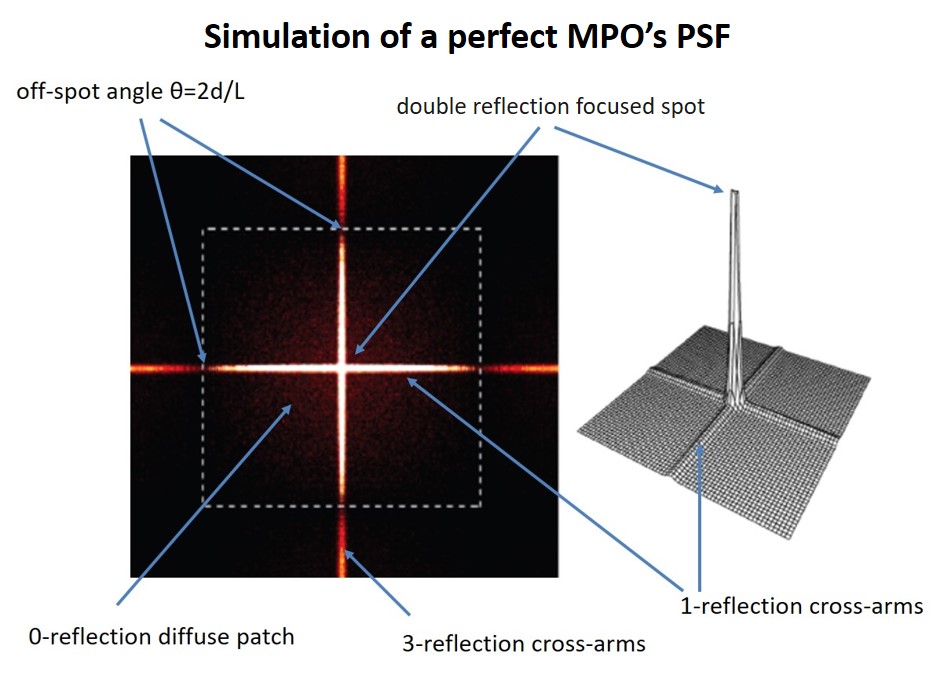}
			\caption{The distinctive PSF created by a slumped MPO \cite{dickspie}. Rays which have undergone two reflections along the pores produce a high intensity central focused spot. Rays which undergo a single reflection in the pores produce the horizontal and vertical cross-arms. Three reflection rays and higher contribute to the outer wings, beyond the first minimum, and the rays which go straight through the MPO create a diffuse background patch.}
	\label{geom2}
\end{figure}

The PSF produced by a single MPO, and an array of MPOs, is very distinctive and unique, as shown in Figure \ref{geom2}. The cross feature formed by these optics comprises a focused spot, horizontal and vertical cross arms and a diffuse patch. The focused spot is created by rays which undergo 2 grazing incidence reflections off orthogonal sides of a single pore, as shown on the left of Figure \ref{geom1}. The vertical and horizontal cross-arms are caused by rays which undergo single or successive odd numbers of reflections off the pore walls, with single reflections off a pore wall creating the cross-arms before the first minima. The arms beyond the first minima are much more faint and are formed by multiple reflections at higher angles. A diffuse patch is created by rays which pass straight-through the MPO, or have undergone multiple even numbers of reflection. Figure \ref{geom2} demonstrates a simulated PSF generated by a single perfect MPO.

For X-ray applications, a classic lobster eye telescope working in the photon energy range 0.2-10 keV, has an optimum $L/d$ ratio (length of pore $L$ and pore width $d$) of \textgreater 50, as the median grazing angle (in radians) for reflection within the pores is given by
\begin{equation}
\theta_{g}=d/L \sim 1.
\end{equation}
Willingale et al.\cite{willf} empirically gives the critical angle of the bare lead glass MPOs as:
\begin{equation}
	\theta_{c}(E)=aE^{-1.04}
	\label{eq:crit}
\end{equation}
where $E$ is the photon energy and $a$=2.4 for $\theta_{c}$ in degrees and $E$ in keV. Equation \ref{eq:crit} demonstrates that the critical reflection angle is a function of the X-ray energy, as the energy increases, the critical angle decreases. This means that the area being used to focus X-rays of both individual MPOs and assembled optics, decreases as the energy increases. This is a significant consideration, as the energy being focused may influence the point spread function if there are variations in the quality of individual MPOs. At C-K (0.28 keV) $\theta_{c}$ is $\sim$9$^{o}$ whereas at Ti-K$_{\alpha}$ (4.51 keV), $\theta_{c}$ is $\sim$0.5$^{o}$, which is very significant, especially for short focal length optics. One example to consider is a 40 mm by 40 mm MPO with 40 $\mu$m pores coated in iridium, 1.2 mm thick and a radius of curvature of 600 mm, assuming an infinite source and the optic is on-axis. At C-K the angle at the edge of the MPO is 1.9$^{o}$ so the whole MPO will contribute to the PSF, but at Ti-K$_{\alpha}$ only the central $\sim$5 mm by 5 mm will. As the corners of the MPO tend to have the poorest form (see Section \ref{prod}) then the on-axis PSF Full Width Half Maximum (FWHM) tends to improve as the energy increases, as the influence of the corners reduces.

The amount of an MPO which contributes to the PSF is also dependent on the length of the channels. If, in the example above, the MPO is 2.4 mm thick, or 1.2 mm thick with 20 $\mu$m channels, then the open angle of the channels is only 0.95$^{o}$. In this case only the central $\sim$10 mm x 10 mm area will provide significant flux in the PSF, as X-rays interacting outside this area will require twice as many reflections to contribute to the PSF. These effects must be taken into consideration when designing the optic for the mission specific energy band.

\subsection{Design of a narrow-field optimised lobster eye telescope}
For wide-field lobster eye designs, $L/d$ is constant across the whole optic aperture, with no specific on-axis point. In comparison, a narrow-field-optimised telescope, which has a varying thickness across the aperture, as described by Angel\cite{ang}, will provide the maximum effective area on-axis. An example of a narrow-field lobster eye optic is the SVOM Microchannel X-ray Telescope (MXT)\cite{gotz}.

Ideally, the full optic would be formed of a continuous series of pores, slumped to a specific radius of curvature, with the length of the pores tapering towards the edges. However, it is not possible to manufacture a single MPO of the required size, nor with a varying thickness. Therefore the optic aperture is formed of smaller tessellated MPOs, with the pore length of each MPO determined by its position in the optic aperture.

To create such an optic, the required pore length (or thickness) for each MPO needs to be calculated, where $r$ is the radial position of the centre of each individual MPO from the centre of the assembled optic aperture.

\begin{equation}
	L = 2 \xi dF/r
	\label{eq:optThick}
\end{equation}
$L$ is calculated using Equation \ref{eq:optThick}. As $F$, the focal length, and $d$, pore diameter, are constant, the thickness of the MPO, or length of the pore, is determined by $r$, the radial position from the centre of the optic, and $\xi$, a scaling factor which is dependent on the energy and the quality (pore alignment, surface roughness etc.) of the MPOs. This indicates that within the optic aperture, the maximum $L/d$ of a single MPO is at the centre of the array, where $r$ is smallest, and reduces towards the outer edge of the assembled optic, where $r$ is largest.

\begin{figure}
	\centering
		\includegraphics[width=0.85\textwidth]{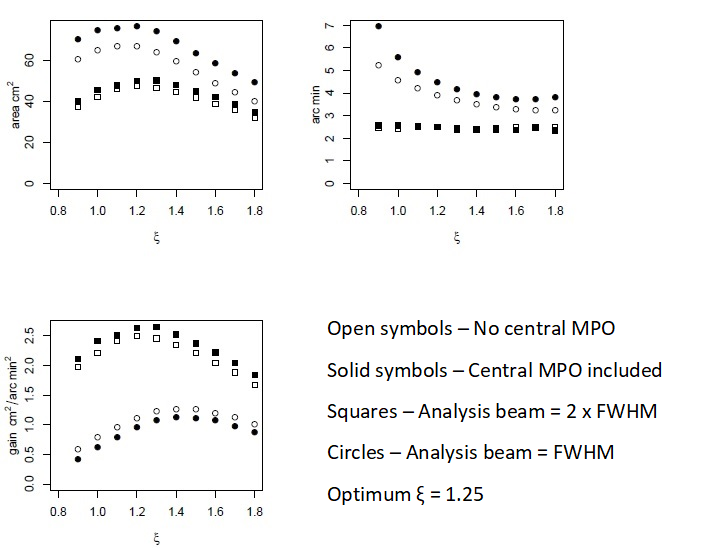}
			\caption{The derivation of the relationship between the MPO's redial position and optimum thickness is $L = 2 \xi dF/r$. $\xi$ is a scaling factor and all simulations were completed at 1.0 keV. The top left pane details the change in on-axis area as a function of $\xi$, which is strongly governed by the thickness of the MPO. The top right pain shows the change in the on-axis FWHM of the double reflection spot given $\xi$ and finally the bottom pane shows the change in gain, the focussing power of the optic, at various values of $\xi$.}
	\label{digr}
\end{figure}

In order to calculate the required thickness, $L$, of each MPO at a given radial position across such an assembled optic, the value of $\xi$ is required. Ray traced simulations were completed at 1 keV to determine the optimum value of $\xi$, assuming 40 $\mu$m pores. The results are shown in Figure \ref{digr}.

The top two panes of Figure \ref{digr} show that if $\xi$ is too low and therefore the MPO is too thin, then the number of double reflections decreases and are replaced by single reflections - reducing the area in the double reflection spot and increasing the FWHM. Similarly, if $\xi$ is large and the MPO is too thick, then the X-rays suffer many more multiple (\textgreater 2) reflections, again reducing the area in the double reflection spot.
Figure \ref{digr} demonstrates how the variation of the value of $\xi$, and therefore the thickness of the MPOs, effects the area, gain and FWHM of the MPO. Here, the gain is a measure of the focusing power of the optic and is the ratio between the total collecting area and the area in the double reflection spot.

As shown in Figure \ref{digr}, the optimum value of $\xi$ at 1 keV is 1.25. This simplifies Equation \ref{eq:optThick}, at 1 keV, to:
\begin{equation}
	L = 2.5dF/r
	\label{eq:optThick2}
\end{equation}

\subsection{Limitations of MPOs}
\label{prod}
If you have a perfect MPO, which is perfectly spherical, with perfectly smooth pores all pointing to exactly the same position on a curved focal plane at exactly the correct focal distance, the minimum PSF size you could achieve would be the width of the pores at that optic-to-detector distance. The angular size of the pore at the focal length of the optic is the fundamental limit of the resolution of an MPO. If every pore is identical and all point to the same position on the focal plane, then the beam from each pore will pileup on top of each other, perfectly, to the width of a pore.

In reality this is not the case and there are many deformations in the form of the MPO which limit the resolution. The full details of the majority of the deformations can be found in Willingale et al.\cite{dickspie}, but they are summarised here.

The three intrinsic aberrations associated with the lobster eye geometry, which limit the angular resolution performance of the optic, independent of the technology used to construct the pore array, are; spherical aberrations, the geometric pore size and
diffraction limits. Non intrinsic aberrations include slumping and formation of the multifibres. Slumping introduces additional radial tilt and shear errors as the pores are stretched and compressed to form the correct profile. Misalignment of multifibres to one another, deformations at the multifibre boundaries and within the multifibres themselves, contribute to the total angular resolution. In addition, the pore surface roughness further increases the angular resolution of the MPO. The combination of the above errors imposes a theoretical limit on the angular resolution of $\sim$2 arcmin for a single MPO, however, the majority of MPOs have an angular resolution far larger than this.

The MPOs are slumped by a technique that sandwiches the MPOs between convex and concave diamond turned mandrels of the appropriate radius of curvature. Equal pressure and heat is applied to both mandrels and across the full surface in order to prevent the shearing of the channels with respect to each other. After slumping, the MPOs and mandrels are left to cool to room temperature which keeps the MPOs form. Unfortunately, trying to slump a square MPO onto a sphere causes deformations in the form of the optic. If you think of trying to wrap a basketball with a square piece of paper, at the centre the fit is very good but towards the corners you get crinkles and folds which distort your piece of paper. This is similar to what happens to an MPO and the end result is that the form of the corner regions of the optics is not as good as at the centre. You can also end up with an astigmatism in the optic where the radius of curvature in one axis does not match that in the other axes. Both of these effects have a massive influence on the net focal length of the optic and the PSF size and shape. At lower energies, the structure of the corners has a strong influence on the PSF but the astigmatism can affect the PSF at all energies.

In addition to the effect of the slumping on individual MPOs, the variation of RoC between the MPOs combined within an assembly will have an affect on the full optic assembly PSF. The optic-to-detector distance of best focus for the optic assembly is governed by the RoC and the form of the frame, but the size of the PSF of the assembly is governed by the individual MPOs. If none of the MPOs have the same RoC as the frame, then they will all be out of focus by varying amounts and this will increase the size of the PSF.

\subsection{Current Missions}
Several missions over the next few years are using this technology in order to take advantage of the large FoV and light-weight nature of these optics for various scientific goals, including planetary science and astronomy. Below is a description of some of the current selected missions.

\subsubsection{BepiColombo}
\label{bep}
\begin{figure}
	\centering
		\includegraphics[width=0.85\textwidth]{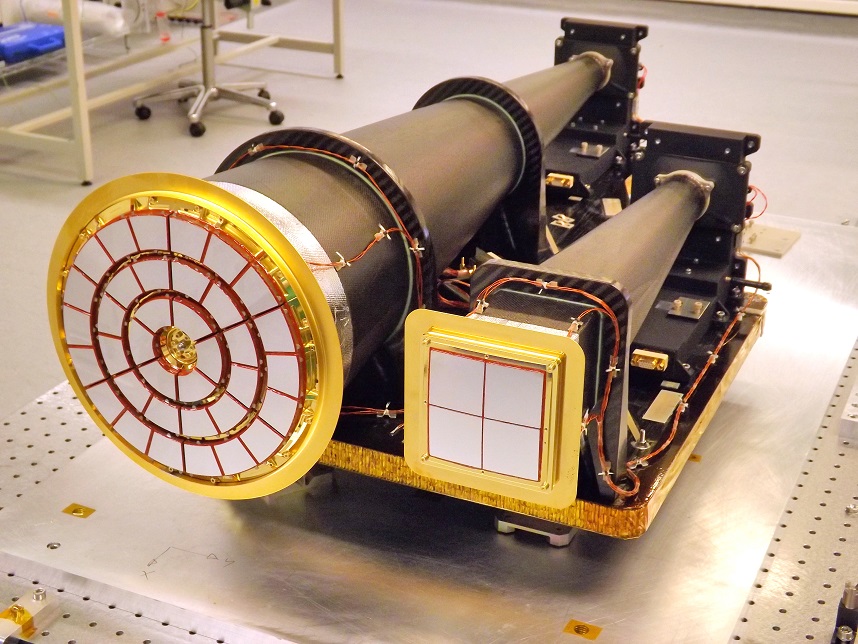}
			\caption{The flight MIXS instrument on the optical bench. The MIXS-T is on the left of the bench, and approximates the Wolter geometry. The MIXS-C is on the right of the bench and is a collimator in a 2 x 2 MPO geometry.}
	\label{bepi}
\end{figure}

The first instrument is the Mercury Imaging X-ray Spectrometer (MIXS)\cite{newmixs} on board the ESA-JAXA mission BepiColombo. Although it was launched in October of 2018, it will not insert into its scientific orbit around Mercury, its destination, until late 2025 - early 2026. MIXS consists of two instruments, the telescope MIXS-T and the wide field collimator MIXS-C, shown on the left and right respectively on the MIXS optical bench in Figure \ref{bepi}. MIXS-T uses the radial packing of 20 $\mu$m square pores and two consecutive sector MPOs, slumped with different radii of curvature to simulate a Wolter geometry\cite{willf}. In order to create the 1 m focal length, the front sectors have a RoC of 4 m and the rear sectors have a RoC of 1.3 m. The FoV of MIXS-T is $\sim$1.1$^{\circ}$ and consists of 36 tandem, sector pairs. The inner ring sectors have a thickness of 2.2 mm, the middle ring optics are 1.3 mm thick and the outer ring optics are just 0.9 mm thick. The MIXS-C instrument uses 20 $\mu$m, square pore, square packed MPOs which are 40 mm by 40 mm in size and 1.2 mm thick. These MPOs have been slumped to a radius of curvature of 550 mm and give a FoV of $\sim$10$^{\circ}$. The complete MIXS instrument on its optical bench weighs $\sim$11 kgs. By using these two instruments side by side, an elemental map of the Mercurian surface using X-ray fluorescence from the solar wind\cite{fras2} will be created.

\subsubsection{SVOM}
\begin{figure}
	\centering
		\includegraphics[width=0.35\textwidth]{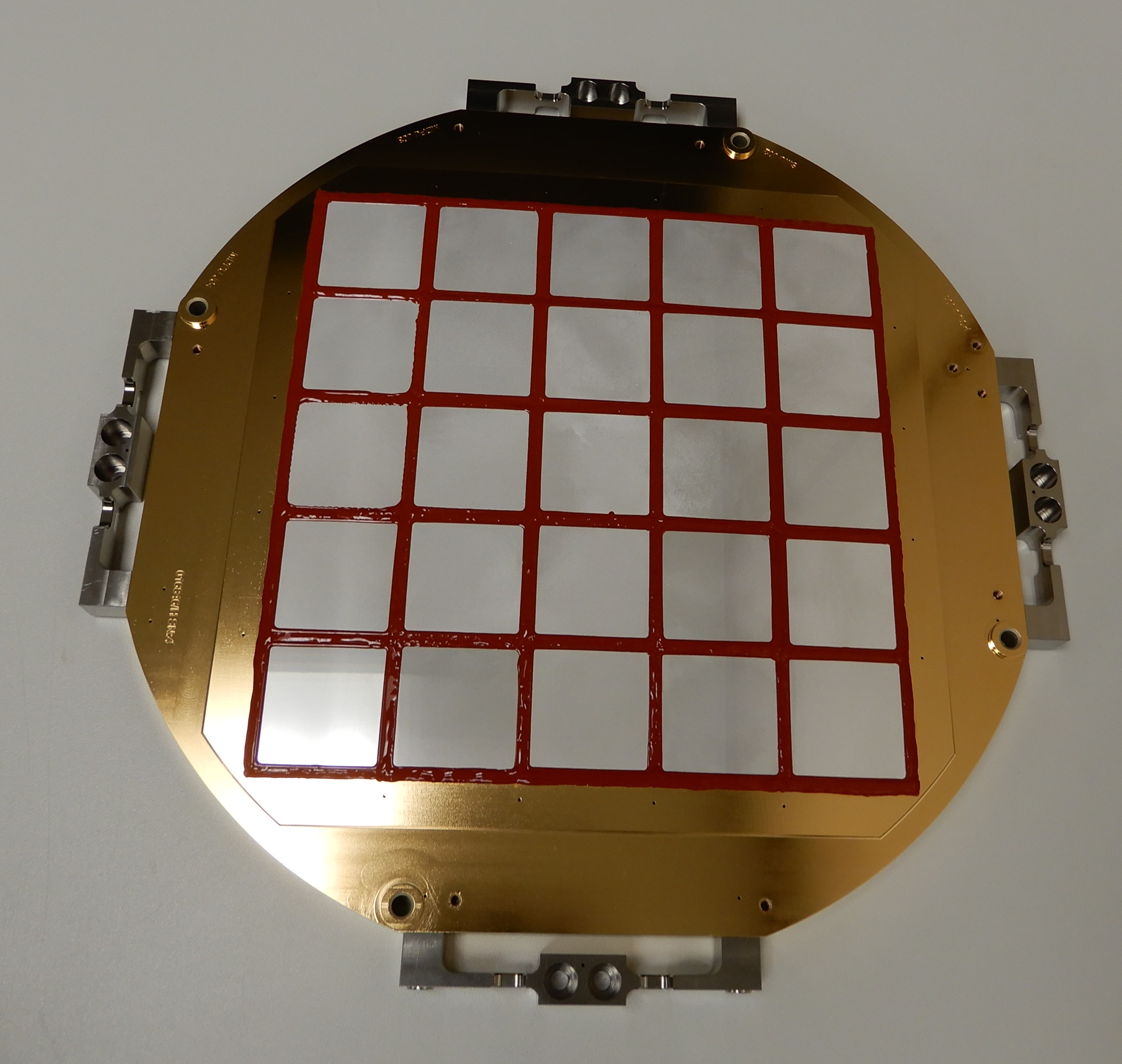}
		\includegraphics[width=0.5\textwidth]{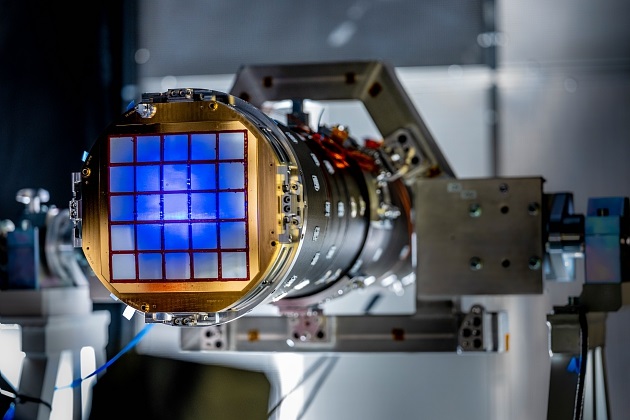}
			\caption{\textit{Left:} the flight MXT optic, a 25 MPO array narrow field lobster eye optic. \textit{Right:} the full flight MXT lobster eye telescope ($\copyright$ T De Prada CNES).}
	\label{mxt}
\end{figure}

The Space-based multi-band astronomical Variable Objects Monitor (SVOM)\cite{kari} is a Chinese – French mission to be launched in 2023. It is comprised of four space borne instruments, including the Microchannel X-ray Telescope (MXT)\cite{gotz}. The MXT’s main goal is to precisely localize, and spectrally characterize X-ray afterglows of GRBs. The MXT is a narrow-field-optimised, lobster eye X-ray focusing telescope, consisting of an array of 25 square MPOs, with a focal length of 1.14 m and working in the energy band 0.2 - 10 keV. The design of the MXT optic (MOP) is optimised to give a 1$^{o}$ detector limited FoV but the optic has the unique characteristics of a lobster eye design, with a wide FoV $>$ 6$^{o}$, and a PSF which is constant over the entire FoV. The MPOs on the Flight Module (FM) MOP have a pore size of 40 $\mu$m giving the optimum thicknesses across the aperture of 2.4 mm in the centre and 1.2 mm at the edges. The left of Figure \ref{mxt} shows the completed FM MOP. Each MPO is 40 mm by 40 mm square and there is a 2 mm gap between each MPO on the frame. The total mass of the fully assembled optic was measured to be 1.43 kg.

\subsubsection{Einstein Probe}
\begin{figure}
	\centering
		\includegraphics[width=0.55\textwidth]{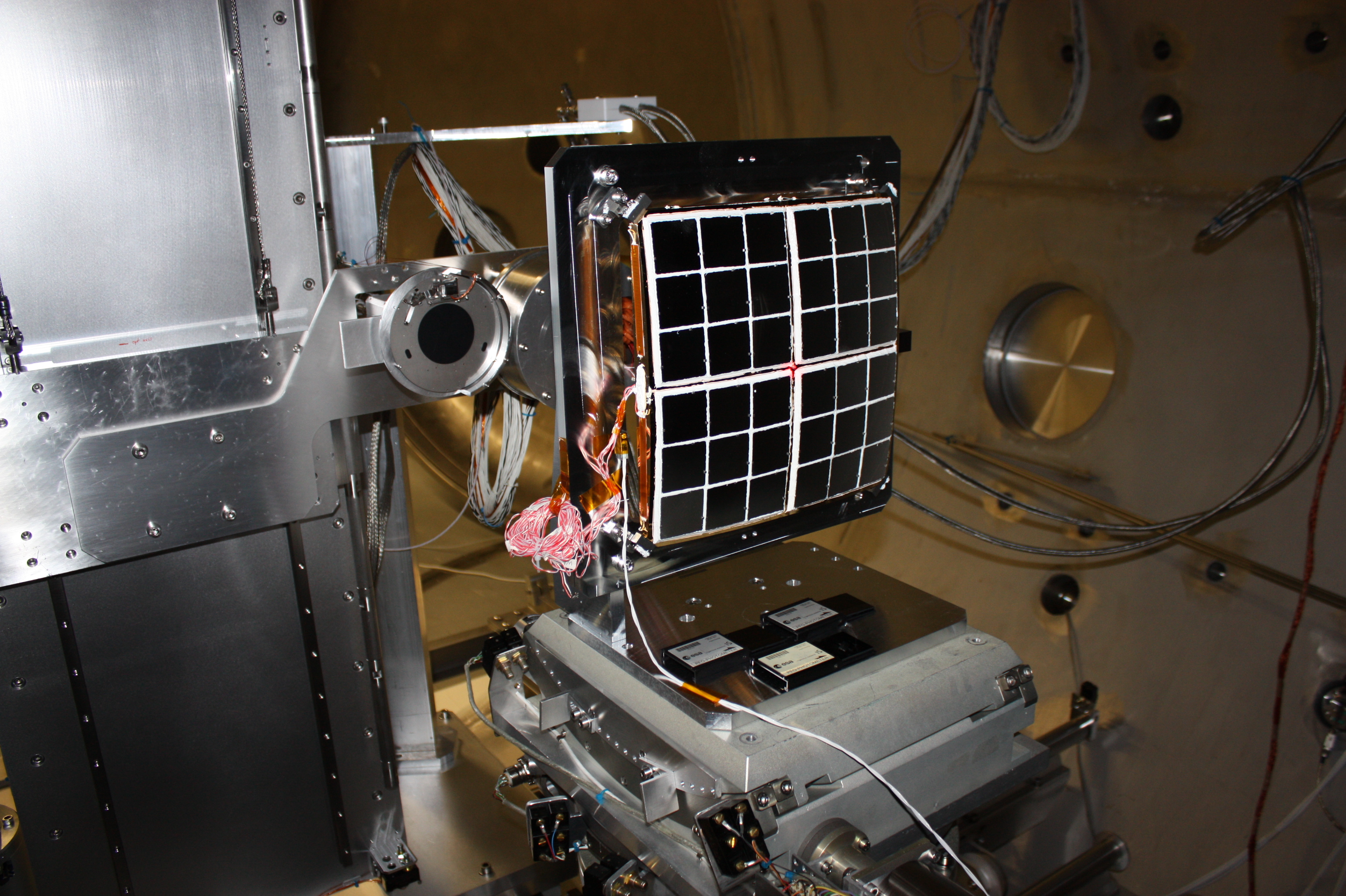}
		\includegraphics[width=0.37\textwidth]{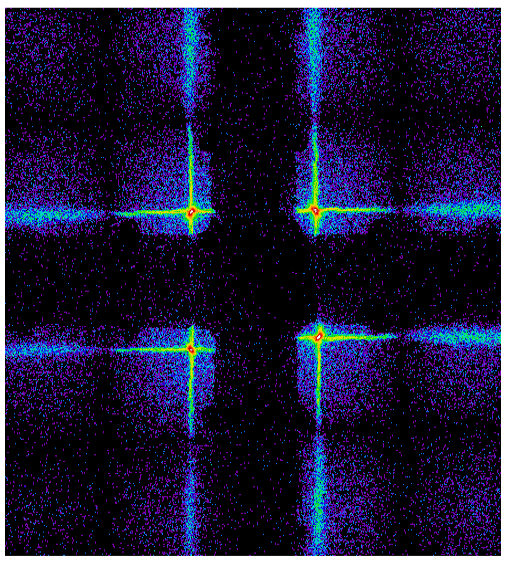}
			\caption{\textit{Left:} A qualification WXT module installed in the PANTER beamline, MPE, Germany, prior to calibration. \textit{Right:} X-ray image with the source centred on the module showing all 4 MPO quadrants focusing. Image taken with Cu-L (0.93 keV) X-rays at PANTER using the TRoPIC camera\cite{tropic}. Images courtesy of MPE.}
	\label{figep}
\end{figure}

Einstein Probe\cite{ep} is a Chinese Academy of Science (CAS) mission due for launch in 2023, with its primary goals to discover high-energy transients and monitor variable objects. The mission consists of two instruments, the Wide field X-ray Telescope (WXT), a lobster eye X-ray telescope consisting of twelve identical modules; and the Follow-up X-ray Telescope (FXT)\cite{fxtspie}, which is a traditional Wolter X-ray telescope. The FXT has been jointly developed by CAS, the European Space Agency (ESA) and the Max Planck Institute for Extraterrestrial Physics (MPE). Each of the WXT modules is comprised of 36 MPOs in a 6 by 6 array (left of Figure \ref{figep}), with a 375 mm focal length, a total FoV of more than 3600 square degrees, an angular resolution goal of 5 arcmin per module and working in the energy range of 0.5-4 keV. Each of the twelve WXT modules, has a focal plane comprised of 4 CMOS detectors in a 2 by 2 array. The modules are aligned so that each 3 by 3 quadrant of MPOs focuses onto a single CMOS detector, thus creating 4 discrete telescopes per module with overlapping FoVs (right of Figure \ref{figep}).

\subsubsection{SMILE}
\begin{figure}
	\centering
		\includegraphics[width=0.85\textwidth]{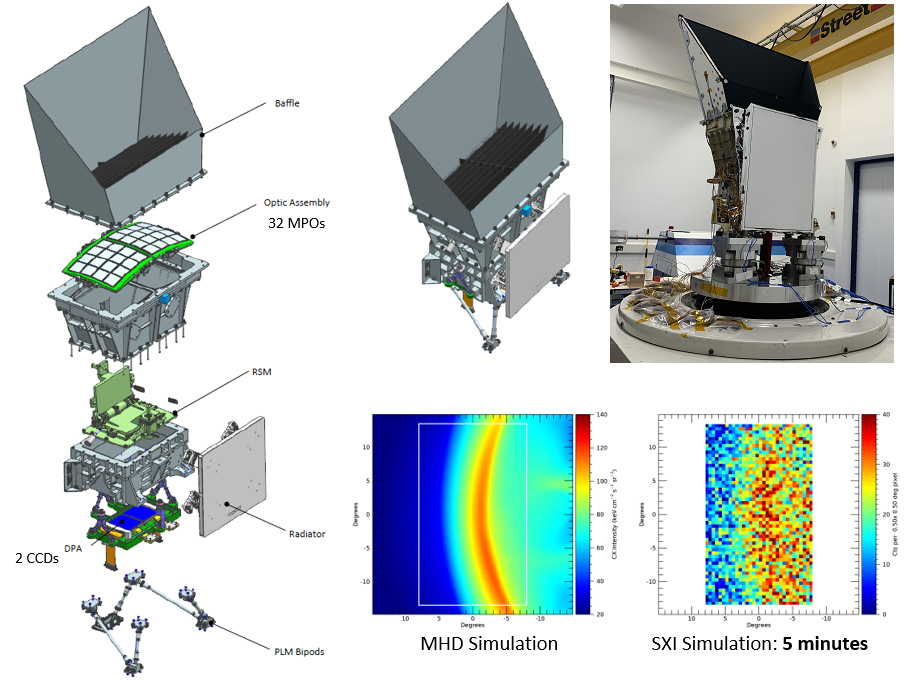}
			\caption{\textit{Clockwise from left:} Exploded CAD diagram of the SXI instrument. CAD diagram of the complete instrument. The structural thermal model of the full instrument during vibration testing. Simulation of a typical event and as seen by SXI after 5 minutes exposure.}
	\label{figsmile}
\end{figure}

Solar wind Magnetosphere Ionosphere Link Explorer (SMILE)\cite{smile} is a joint mission between ESA and CAS to investigate the dynamic response of the Earth's magnetosphere to the impact of the solar wind. From an elliptical polar orbit it will combine soft X-ray imaging of the Earth's magnetopause and magnetospheric cusps with simultaneous UV imaging of the Northern aurora, and will monitor in situ the solar wind and magnetosheath plasma conditions so as to set the imaging data into context. It is due for launch in late 2024 or early 2025 with 4 separate instruments on board, including the Soft X-ray Imager (SXI). The SXI is an elongated lobster eye telescope with an array of 4 by 8 MPOs. Each MPO is 40 mm by 40 mm, with iridium coated 40 $\mu$m pores and a focal length of 300 mm. The high charge state solar wind ions in collision with hydrogen produce photons at soft X-ray (and EUV) energies within the 0.2 keV to 2.5 keV band. The focal plane consists of 2 CCDs and the instrument has a FoV of 26.5$^{o}$ by 15.5$^{o}$. The wide FoV enables SXI to spectrally map the location, shape, and motion of Earth's magnetospheric boundaries. Figure \ref{figsmile} shows an exploded CAD diagram of the SXI instrument on the left, the structural thermal model of the full instrument during vibration testing on the top right, and a simulation of the data expected on the bottom right.

\section{Lobster eye optics in MFO/Schmidt arrangement}
\label{secsch}

\subsection{Schmidt objectives}
The lobster eye geometry X-ray optics offer an excellent opportunity to achieve very wide fields of view. One dimensional lobster eye geometry was originally suggested by Schmidt \cite{Schmidt}, based upon flat reflectors. The device consists of a set of flat reflecting surfaces. The plane reflectors are arranged in an uniform radial pattern around the perimeter of a cylinder of radius R. X-rays from a given direction are focused to a line on the surface of a cylinder of radius R/2 (Fig. 2). The azimuthal angle is determined directly from the centroid of the focused image.  At glancing angle of X-rays of wavelength 1 nm and longer, this device can be used for the focusing of a sizable portion of an intercepted beam of parallel incident X-rays. Focusing is not perfect and the image size is finite. On the other hand, this type of focusing device offers a wide FoV, of up to a maximum of the half sphere of the coded aperture. It is possible to achieve an angular resolution on the order of one tenth of a degree or better. Two such systems in sequence, with orthogonal stacks of reflectors, form a double-focusing device. Such a device offers a FoV of up to 1000 square degrees at a moderate angular resolution.

\begin{figure}
\begin{center}
\includegraphics[width=45mm]{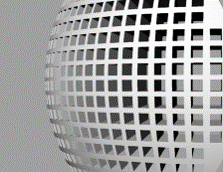}
\includegraphics[width=36mm]{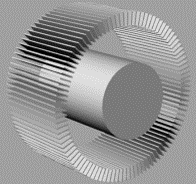}
\includegraphics[width=33mm]{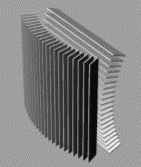}
\caption{The arrangements of Angel MPO (left), Schmidt MFO 1D (middle) and Schmidt 2D lobster eye optics\cite{Sve03}.}
\label{Fig2}
\end{center}
\end{figure}

\begin{figure}
\begin{center}
\includegraphics[width=\columnwidth]{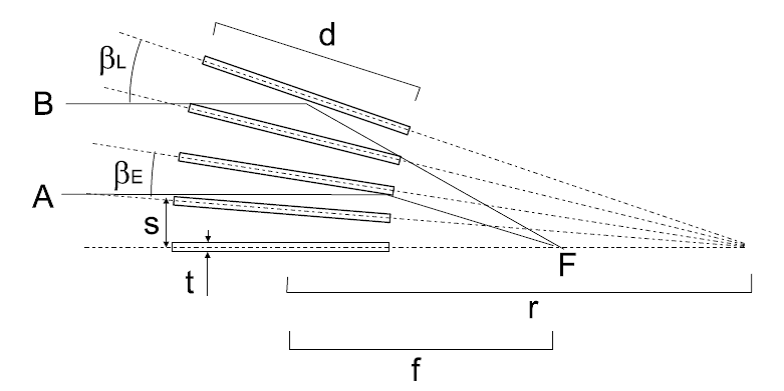}
\caption{The schematic arrangement of the Schmidt lobster eye type X-ray optics used for simple equations derivation \cite{Sve03}.}
\label{Fig2}
\end{center}
\end{figure}

 It is obvious that this type of wide-field X-ray telescope could play an important role in future X-ray astrophysics. These innovative very wide field X-ray telescopes have only recently been suggested for space-based applications. One of the first proposals was the All Sky Supernova and Transient Explorer (ASTRE, Gorenstein \cite{Goren79}\cite{Goren87}).  This proposal included a cylindrical coded aperture outside of the reflectors, which provide angular resolution along the cylinder axis. The coded aperture contains circumferential open slits that are 1 mm  wide and are in a pseudo-random pattern. The line image is modulated along its length by the coded aperture. The image is cross-correlated with the coded aperture to determine the polar angle of one or more sources. The FoV of this system can be, in principle, up to 360$^{\circ}$ in the azimuthal direction and nearly 90\%  of the solid angle in the polar direction. To create this mirror system, a development of double-sided flats is necessary.  There is also potential for extending the wide field imaging system to higher energy with the application of multi-layers or other coatings in analogy to those described for flat reflectors in the K-B geometry. 
 
 The angular resolution of the lobster eye optics in the Schmidt arrangement is a function of  spacing between the reflecting plates and focal length. In the Schmidt arrangement, the lobster eye consists of plates of thickness $t$, and depth $d$ (Fig. 12). Spacing between plate planes is $s$, focal length $f$, radius $r$, focal point $F$, and $\beta$ is the angle between optical axis and focused photons. Beam A (Fig. 12) shows the situation where the plate is fully illuminated and the crossection of the plate is maximal (effective reflection).  Beam B is the last beam that can be reflected into the focal point. Beams that are further from the optical axis reflect more than once (critical reflection). If reflected twice from the same set of plates, the photon does not reach the focal point and continues parallel to the incoming photon direction \cite{Sve03}.

 If $t \ll s \ll d \ll f$ we can derive the following simple equations\cite{Sve03} \cite{Inne01}, where $\alpha$ is the estimate of the angular resolution.
 \begin{center}
 
$f =\frac{r}{2}$

$\beta _E = \frac{(s-t)}{d}$

$\beta _L = 2 \beta _E$    

$\alpha \sim\frac{2s}{r} =\frac{s}{f}$


\end{center}

 
 The design concept is different for lobster eye systems based on two reflections, 
a single reflection on a horizontally oriented surface (pore wall or mirror)
and a single reflection on a horizontally oriented surface.
Particularly, this is a case of Schmidt lobster eye.
A paper by Tichy et al. \cite{Tichy2019} presents analytical formula allowing direct computing
of the effective collecting area for those systems by the formula
\begin{equation}
L(r,s,t,\zeta) = 2r \frac{s}{s+t} \frac{
  {\widetilde{\cal R}}(2 \zeta)
  - 2 {\widetilde{\cal R}}(\zeta)
  + {\widetilde{R}}(0)
  }
  { \zeta},
\end{equation}
where ${\widetilde{\cal R}}(\theta):= \int \int {\cal R}(\theta) \mbox{d} \theta \mbox{d} \theta
=\int {\bar{\cal R}}(\theta) \mbox{d} \theta $
is an arbitrary second antiderivative of $\cal R$.
Radius of the system measured to mirror center is denoted $r$,
$s$ represents mirror spacing (or pore width)
and $t$ is mirror (pore wall) thickness.
The effective collecting area equals $L^2$ for the Angel system
and $L_1L_2$ for the Schmidt system, where $L_1$ and $L_2$ are related to individual 
mirror stacks as they have different radii and they may differ in other parameters.
The value $\zeta$ is the  ratio between mirror (pore) depth $d$ and $s$.
The optimal value of this ratio is given by the reflectivity function for given surface and photon energy only. 
The paper by Tichy et al. \cite{Tichy10b} presents the detailed procedure for how the optimal value of this ratio
can be analytically calculated.

In addition, a paper by Tichy and Willingale \cite{Tichy2018} presents a formula for the optimal value of $s$ as
\begin{equation}
  s=-t+ \sqrt{2Rt \zeta +t^2}
\end{equation} 
Here, $R=(r+d/2)$ is the radius of the system measured to the front aperture (d is the mirror depth).
This solves a common problem when focal length is limited e.g. by available
space in a spacecraft. Mirror (pore wall) thickness $t$ should be as small as possible but must be large enough to achieve sufficient stress endurance, etc.

\begin{figure}
\begin{center}
\includegraphics[width=50mm]{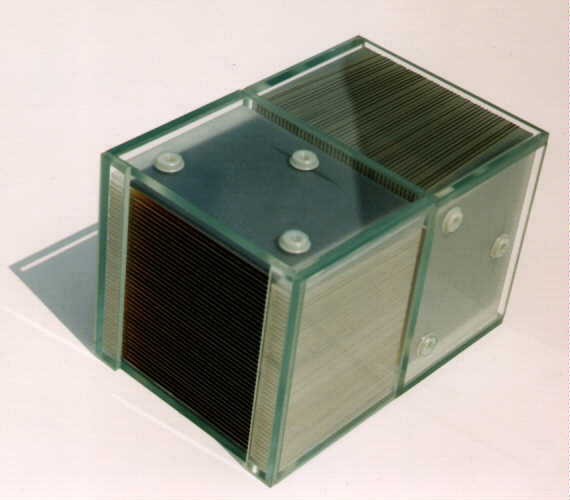}
\includegraphics[width=60mm]{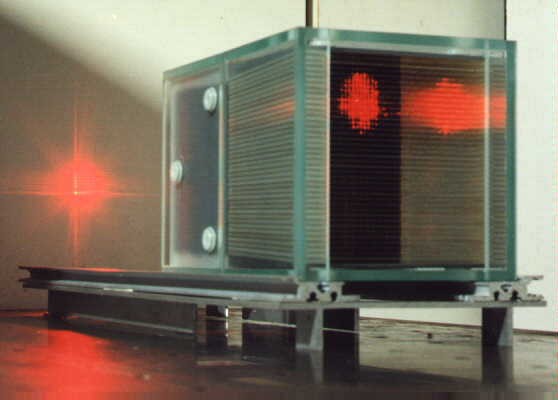}
\caption{The Schmidt objective midi test module with 100 mm x 80 mm plates (left) and its optical tests (right).}\label{Fig2}
\end{center}
\end{figure}

\begin{figure}
\begin{center}
\includegraphics[width=50mm]{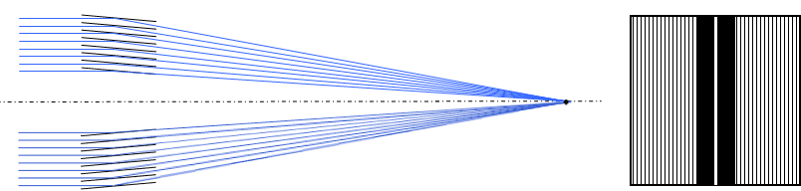}
\includegraphics[width=50mm]{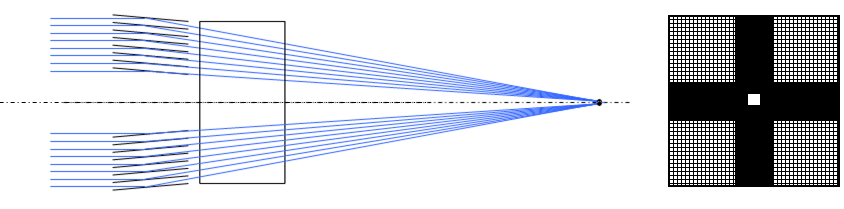}
\caption{The schema of Schmidt lobster eye modules, 1D (left) and 2D (right) arrangements.}\label{Fig2}
\end{center}
\end{figure}

The 1D and 2D lobster eye Schmidt modules are illustrated in Figs. 14 and 17.
To test the design and assembly of lobster eye modules in Schmidt geometry, various test modules were manufactured and tested (Table 1).
 
 \begin{figure}
\begin{center}
\includegraphics[width=60mm]{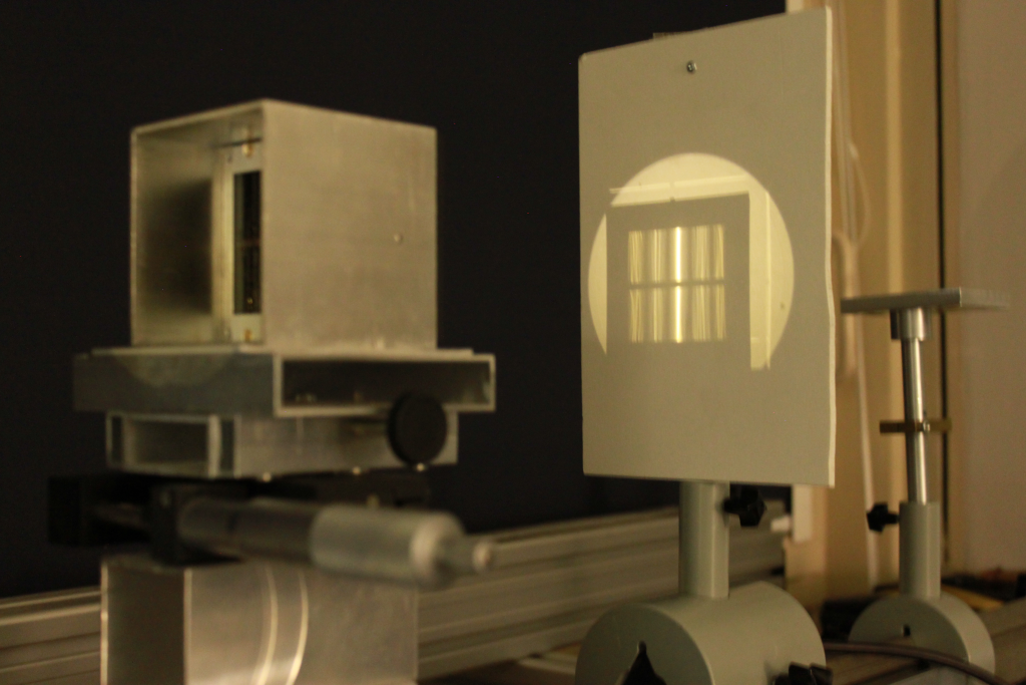}
\includegraphics[width=52mm]{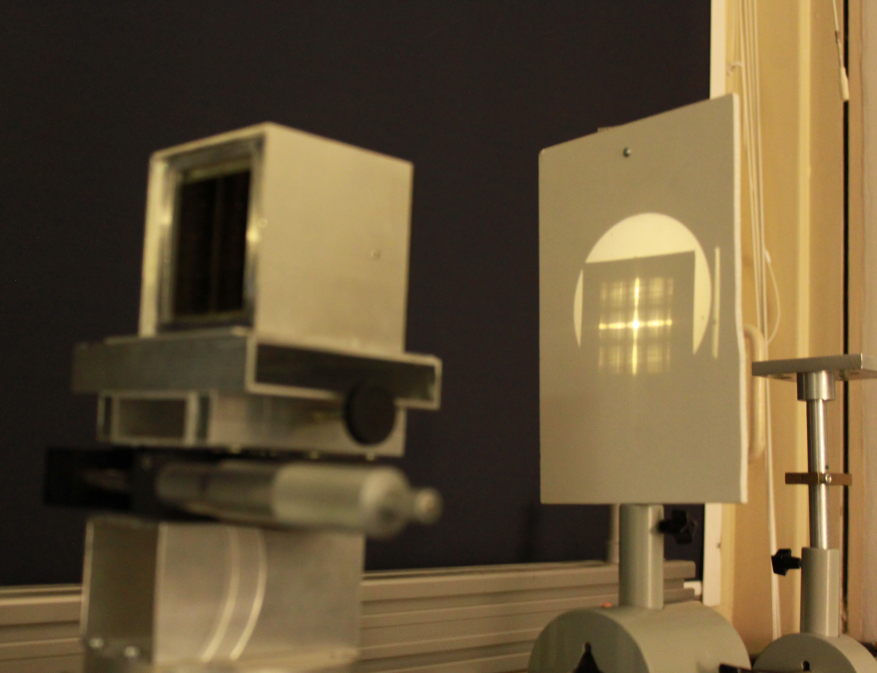}
\caption{The visual tests for the 1D (left) and 2D lobster eye modules (right) in Schmidt arrangements.}\label{Fig2}
\end{center}
\end{figure}

 The first lobster eye X-ray Schmidt telescope prototype (midi) consisted of two perpendicular arrays of flats (36 and 42 double-sided flats 100 mm x 80 mm each). The flats were 0.3 mm thick and gold-coated. The focal distance was 400 mm from the midplane. The FoV was about 6.5 degrees (Fig. 13). The results of optical and X-ray tests indicated a performance close to those provided by mathematical modelling (ray-tracing). X-ray testing was carried out in the test facility of the X-ray astronomy group at the University of Leicester.  At a later date, test modules with a Schmidt geometry were designed and developed using 0.1 mm thick gold coated glass plates that were 23 mm x 23 mm, with a 0.3 mm spacing. The aperture/length ratio is 80. A single module has 60 plates. Two analogous modules represent the 2D system for laboratory tests, providing focus to focus imaging with focal distances of 85 and 95 cm. The innovative gold coating technique resulted in a final surface micro roughness rms to 0.2-0.5 nm. Various modifications of this arrangement have been designed both for imaging sources at final distances (for laboratory tests) as well as for distant sources (the corresponding double--focusing array has f = 250 mm and FoV = 2.5 deg). In parallel, numerous ray-tracing simulations have been performed, allowing for a comparison between theoretical and experimental results.

\begin{figure}
\begin{center}
\includegraphics[width=55mm]{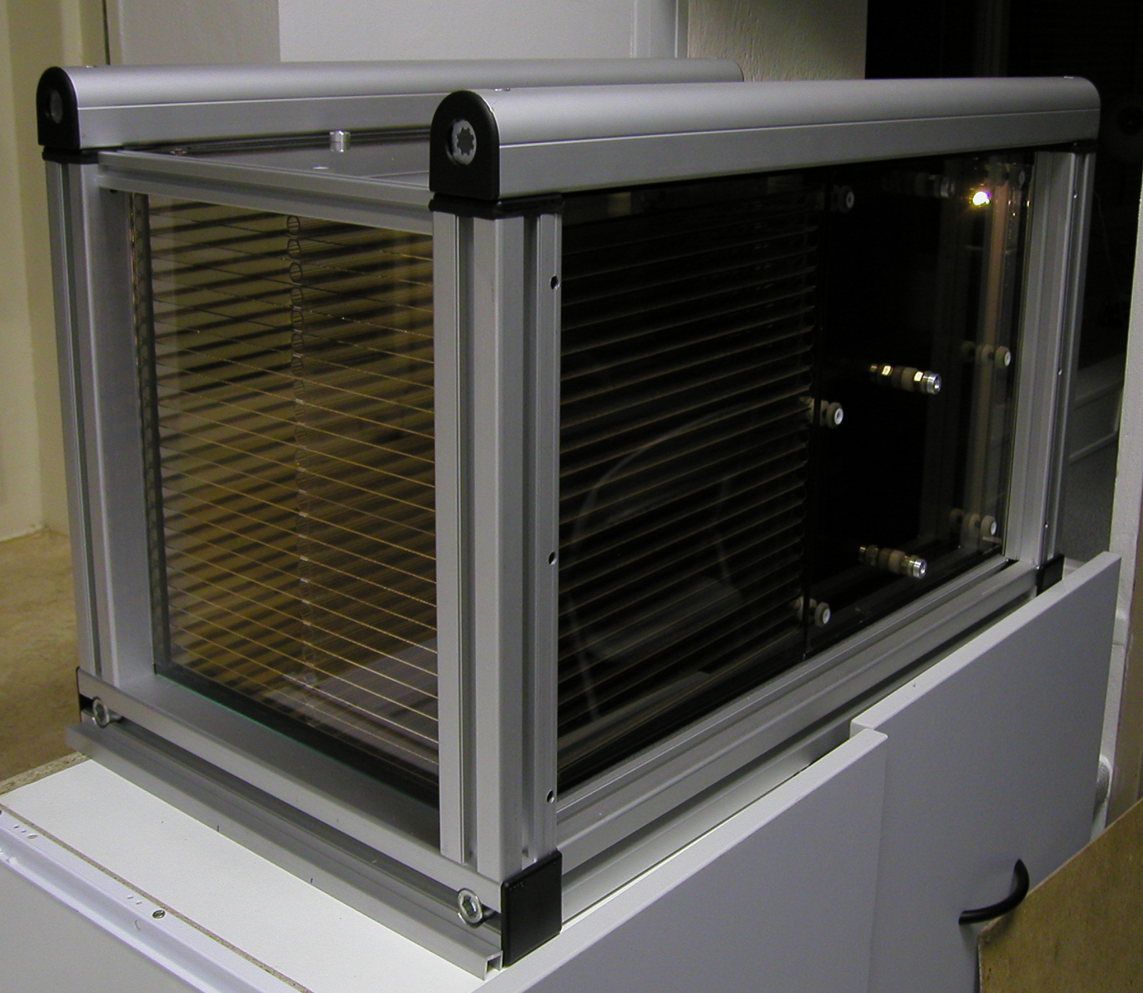}
\includegraphics[width=51mm]{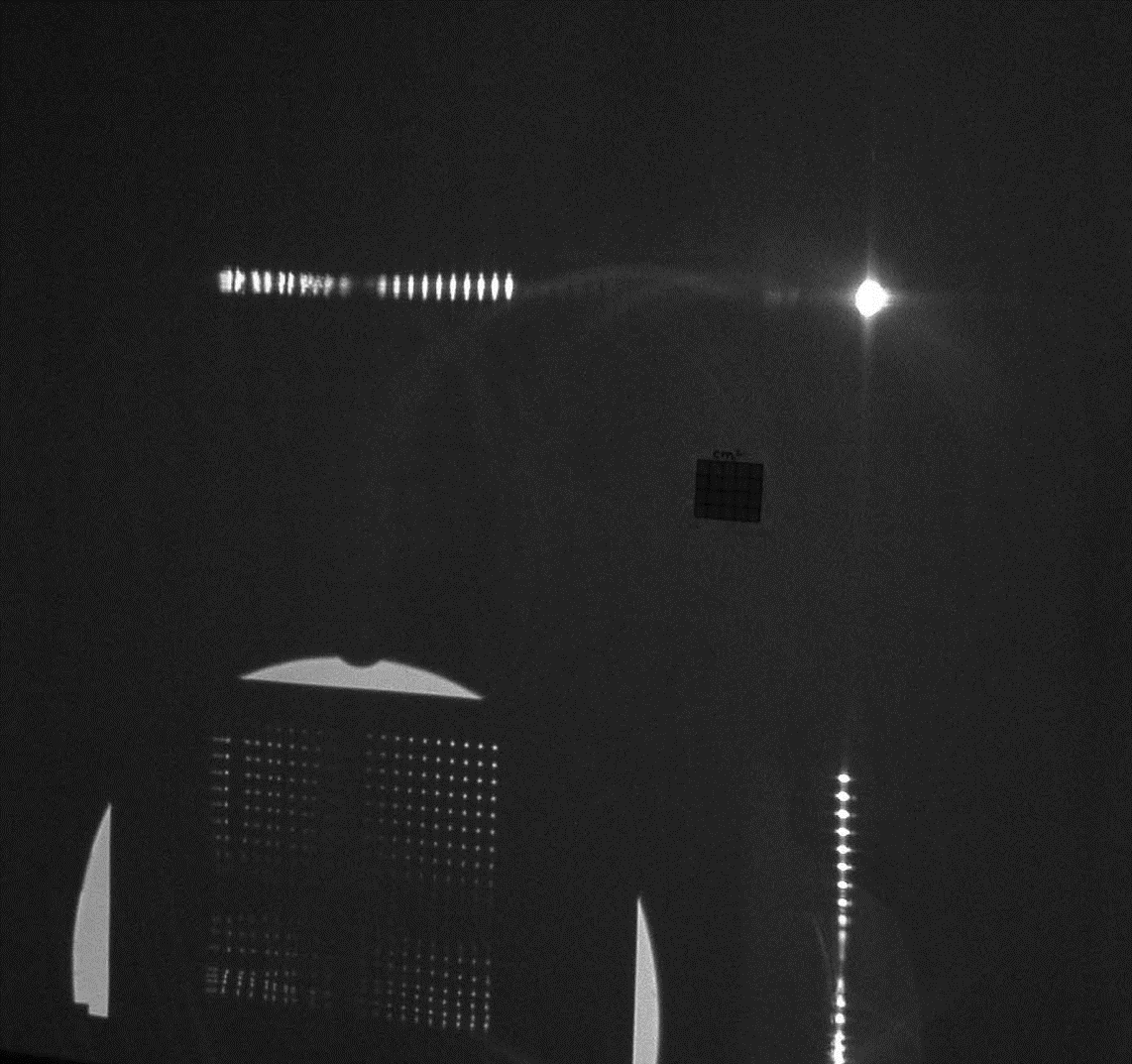}
\caption{The large Schmidt lobster eye module macro with aperture of 30 x 30 cm (left) and its optical image in visible light (right).}
\label{Fig2}
\end{center}
\end{figure}

\begin{figure}
\begin{center}
\includegraphics[width=60mm]{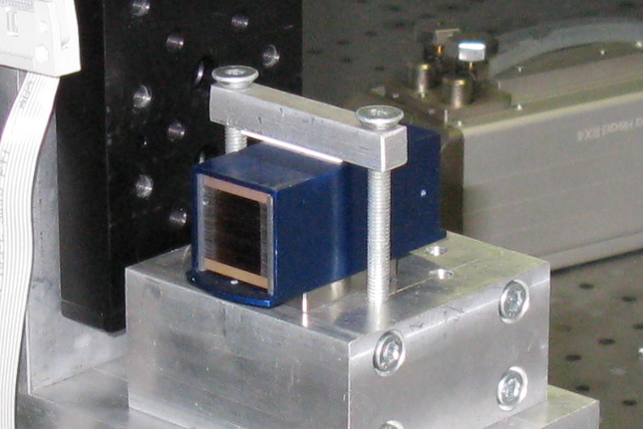}
\includegraphics[width=53mm]{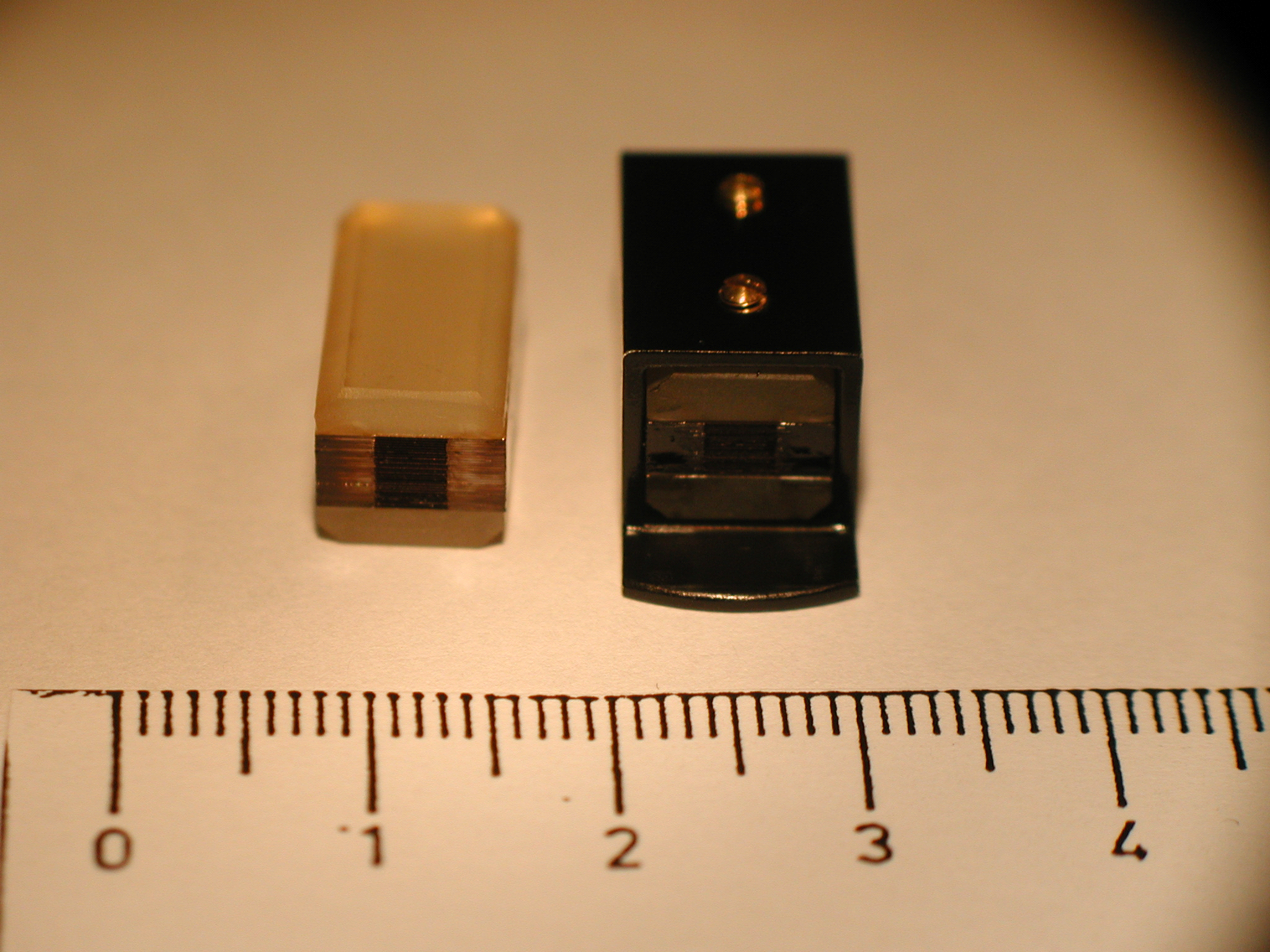}
\caption{The LE Mini and LE Micro modules.}\label{Fig2}
\end{center}
\end{figure}


\begin{table}[h!]
\centering
\begin{tabular}{||c c c c c c c c c c||} 
 \hline
 Module & Size  & Plate thickness & Distance & Length  & Eff. angle & f  & Resolution  & FoV  & Energy \\ [0.5ex] 
 \hline\hline
 Macro & 300 & 0.75 & 10.8 & 300 & 0.036 & 6 000 & 7 & 16 & 3 \\ 
 Middle & 80 & 0.3 & 2 & 80 & 0.025 & 400 & 20 & 12 & 2 \\
 Mini 1 & 24 & 0.1 & 0.3 & 30 & 0.01 & 900 & 2 & 5 & 5 \\
 Mini 2 & 24 & 0.1 & 0.3 & 30 & 0.01  & 250   & 6  & 5 & 5\\
 Micro & 3 & 0.03 & 0.07 &  14  & 0.005  & 80 & 4  & 3 & 10  \\ [1ex] 
 \hline
\end{tabular}
\caption{The parameters of selected test Schmidt lobster eye modules assembled and tested. The distance parameter means the separation between reflecting foils. The parameters size, plate thickness, distance, length and focal distance f are given in mm, resolution in arcmin, FoV in degrees and optimal energy in keV.}
\label{table:1}
\end{table}

\begin{figure}
\begin{center}
\includegraphics [width=100mm]{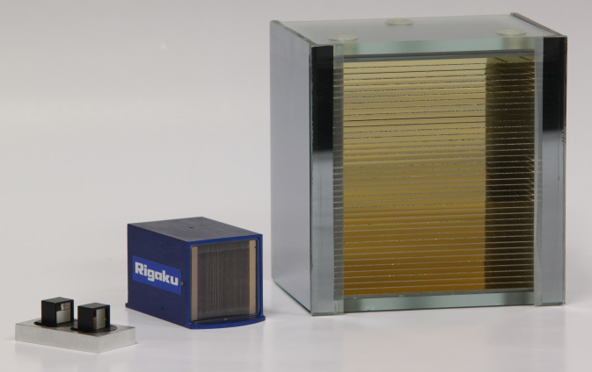}
\caption{The various sizes of Schmidt MFO lobster modules manufactured by Rigaku Prague (Photo courtesy Rigaku Prague).}
\label{Fig2}
\end{center}
\end{figure}

\begin{figure}
\begin{center}
\includegraphics[width=100mm]{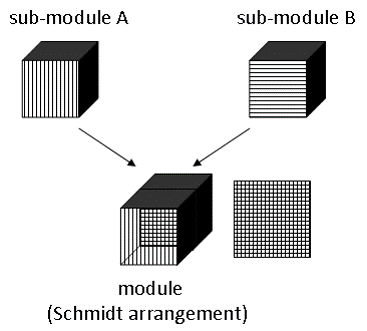}
\caption{The schematic assembly of 1D lobster eye Schmidt sub-modules and 2D lobster modules.}\label{Fig2}
\end{center}
\end{figure}

\begin{figure}
\begin{center}
\includegraphics[width=\columnwidth]{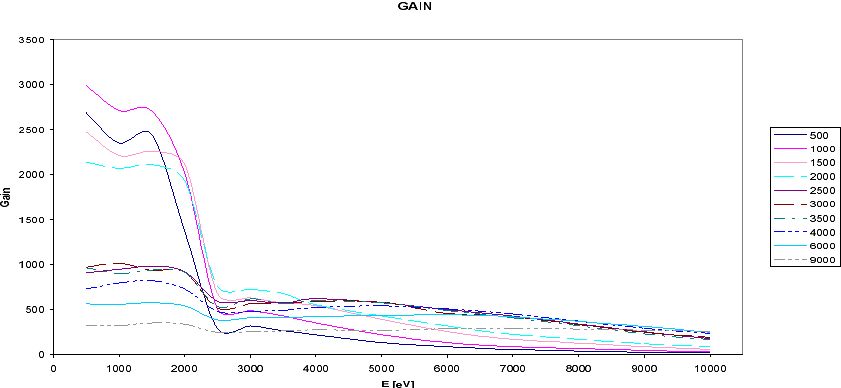}
\caption{The calculated on axis gain dependence on energy for lobster eye modules in Schmidt geometry. The f=375 mm, gold coated plates 100 microns thick \cite{Sve03}. 
}\label{Fig2}
\end{center}
\end{figure}

\begin{figure}
\begin{center}
\includegraphics[width=40mm]{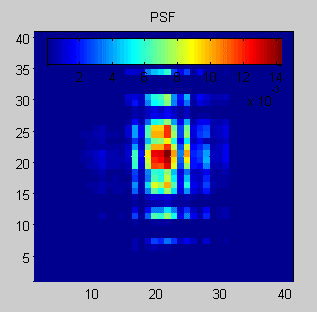}
\includegraphics[width=60mm]{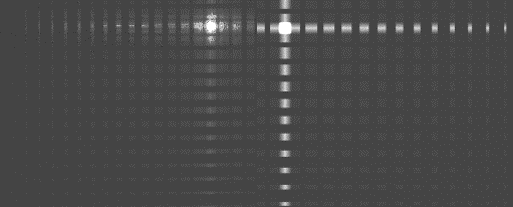}
\caption{The point-to-point focusing system, lobster eye Schmidt micro (3x3x0.03 mm mirrors), source-detector distance 160 mm, 8 keV photons, left, X-ray experiment vs. simulation, point-to-point focusing system, lobster eye Schmidt mini (25x25x0.1 mm , source-detector distance 1.2 m, 8 keV photons, image width: 2x512 pixels, 24 micron pixel, Gain: 570 (measured) vs. 584 (model)
(right).}
\label{Fig2}
\end{center}
\end{figure}

Following the aforementioned developments, even smaller (micro) lobster eye modules were constructed and tested in both visible light and X-rays. As an example, we show X-ray test results for the mini and micro lobster eye modules (Fig. 21).  These results show the on-axis and off-axis imaging performance of the lobster eye module tested. For mosaics of X-ray test images for various energies see Fig. 22 and for various off-axis angles at 4.5 keV see Fig. 23.

\begin{figure}
\begin{center}
\includegraphics[width=100mm]{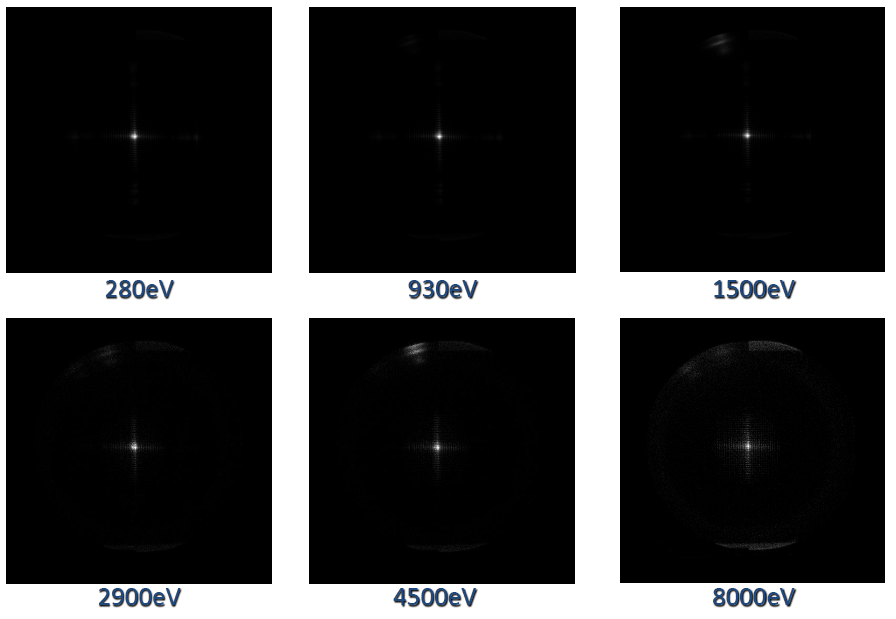}
\caption{X-ray images for various energies from MFO Schmit lobster eye mini, f=25 cm, Palermo X-ray test facility \cite{Tichy10b}}
\label{Fig2}
\end{center}
\end{figure}

\begin{figure}
\begin{center}
\includegraphics[width=120mm]{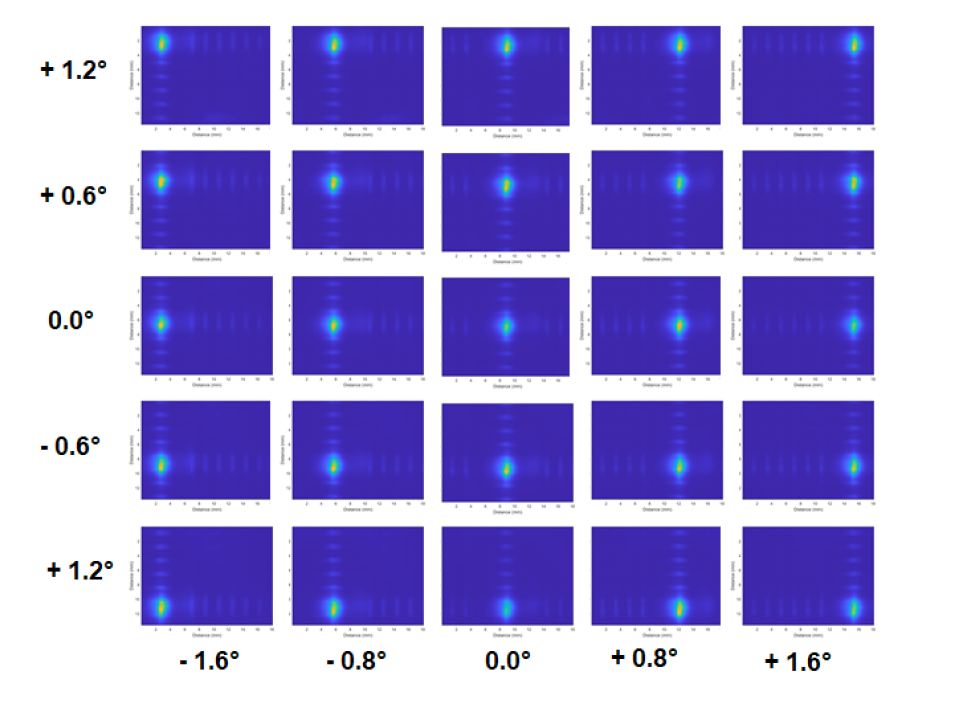}
\caption{Mosaic of X-ray (at 4.5 keV) images for various off-axis positions (demonstrating the off-axis imaging performance), lobster eye REX2 2D module (83 gold coated glass foils 148 x 57 x 0.42 mm each, FoV 4.7 x 4.3 deg.)  \cite{Pina21}}
\label{Fig2}
\end{center}
\end{figure}


\subsubsection{Substrates for lobster eye lenses in Schmidt/MFO arrangement}

In general, there is growing need for large segmented X-ray foil telescopes of various geometries and geometrical arrangements. The requirement of minimizing the weight of future large X-ray space telescopes and at the same time achieving large collecting 
area for future large astronomical telescopes can be met with thin X-ray-reflecting foils (i.e., thin, lightweight, multiple layers that can be easily nested to form precise high-throughput mirror assemblies). This includes the large modules of the Wolter 1 geometry, the large Kirkpatrick-Baez (further referred as K-B) modules (as they can play an important role in future X-ray astronomy projects as a promising and less laborious to produce alternative) as well as the large lobster eye modules in the Schmidt arrangements. Although these particular X-ray optics modules differ in the geometry of foils/shells arrangements, they do not differ much from the point of the view of the foils/shells production and assembly, and also share all the problems of calculations, design, development, weight constraints, manufacture, assembling, testing, etc. It is evident that these problems are common and rather important for the majority of the large aperture X-ray astronomy space-based observatories. Most of  the future space projects require light material alternatives \cite{Hud2011}. 

We (Czech team with participation of the first author of this chapter) have developed  various prototypes of the above mentioned X-ray optics modules based on high quality X-ray reflecting gold coated float glass foils \cite{Hud00}. The glass represents a promising alternative to electroformed nickel shells used in Wolter optics, the main advantage being much lower specific weight (typically 2.2 g cm$^{-3}$ if compared with 8.8 g cm$^{-3}$ for nickel). For the large prototype modules of dimensions equal to or exceeding 30 x 30 x 30 cm, mostly glass foils of thickness of  0.75 mm have been used, although in the future this thickness can be further reduced down to 0.3 mm and perhaps even less (we have successfully designed, developed and tested systems based on glass foils as thin as 30 microns, albeit for much smaller sizes of the modules). More recently, Silicon wafers with superior flatness and micro-roughness are serving  as alternative substrates for lobster eye MFO modules. The recent HORUS experiment can serve as an example. HORUS has 4 modules,
2 modules with Au surface,  
2 modules with  Ir surface,
each module has 17 silicon foils, i.e. in total 4 x 17 Si wafers 0.625 mm thick, with an aperture of 85 x 65 mm f=2 m. The goal
is to experimentally compare different reflective layers (Fig. 24).

\begin{figure}
\begin{center}
\includegraphics[width=100mm]{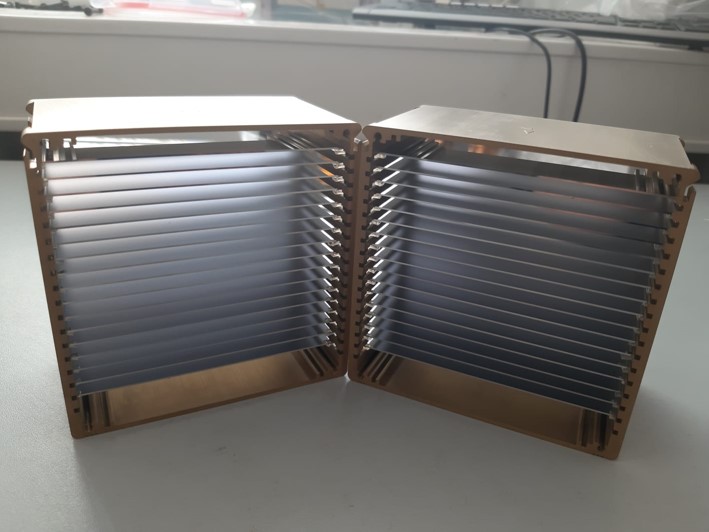}
\caption{The experimental HORUS modules with Si wafers \cite{Stehlik21}}
\label{Fig2}
\end{center}
\end{figure}

These substrates, both glass foils and silicon wafers, can be used in various X-ray optics arrangements using MFO technology, mostly lobster eye and K-B \cite{Hud2009}\cite{Hud2016c}.


\subsubsection{The application and the future of lobster eye telescopes in Schmidt arrangements}

It is obvious that the lobster eye Schmidt MFO prototypes confirm the feasibility to design and develop these telescopes with currently available technologies. Considerations for fabricating and assembling a wide-field space-based X-ray observatory include: (1) Reduction of the micro-roughness and slope errors of the reflecting surfaces to optimize the angular resolution and reflectivity/efficiency of the system. The past development has already led to significant micro roughness improvement (to 0.2-0.5 nm for glass substrates and 0.1 nm for silicon substrates) (2) The design and construction of larger or multiple modules to achieve a larger FoV (of order of 1000 square degrees and/or more) and enhance the collecting area (3) Reduction in the spacing and plate thickness (Schmidt arrangement) to improve imaging performance (angular resolution and system efficiency) and (4) Advanced, alternative layer applications, and/or other approaches applied to the reflecting surfaces to improve the reflectivity and to extend the energy bandpass to higher energies.

The application of very wide field Schmidt MFO X-ray imaging systems could be without doubt very valuable in many areas of X-ray and gamma-ray astrophysics. Results of analyses and simulations of lobster eye X-ray telescopes have indicated that they will be able to monitor the X-ray sky at an unprecedented level of sensitivity, an order of magnitude better than any previous X-ray all-sky monitor. Limits as faint as 10$^{-12}$ erg cm$^{-2}$ s$^{-1}$ for daily observation in the soft X-ray range (typically 1-10 keV) are expected to be achieved, allowing monitoring of all classes of X-ray sources, including X-ray binaries, fainter classes such as AGNs, coronal sources, cataclysmic variables, as well as fast X-ray transients including GRBs and the nearby Type II supernovae\cite{Hud2012}\cite{Hud2018c}\cite{Hudec2006c} . For pointed observations, limits better than 10$^{-14}$ erg cm$^{-2}$ s$^{-1}$ (0.5 to 3 keV) could be obtained, sufficient enough to detect X-ray afterglows to GRBs \cite{Sve04}\cite{Hud2013}. 

\subsubsection{Lobster eye Laboratory Modifications} 

The lobster eye soft X-ray optics, originally proposed and designed for astronomical (space) applications, has  potential for numerous laboratory applications.

 As an example, lobster eye optics can be modified for efficient collection of laser-plasma radiation for wavelengths longer than 8 nm \cite{Bart}. The optics for this application  consist of two orthogonal stacks of ellipsoidal mirrors forming a double-focusing device \cite{Bart}. The ellipsoidal surfaces were covered by a layer of gold that has relatively high reflectivity at the wavelength range that is 8-20 nm up to an incident angle of around 10 degrees. 
 The width of the mirrors forming the optics assemblies is 40 mm. As can be noticed, the spacing between adjacent mirrors increases with the distance from the axis. The curvature of the mirrors and the spacing between them were optimized using ray tracing simulations to maximize the optics aperture and to minimize the size of the focal spot. 


\subsubsection{Hybrid lobster eye}

The lobster eye Schmidt MFO configuration described in the previous sections is a 
wide-field, relatively low angular resolution optics. Achieving finer angular resolution is challenging given the current limitations of the technological limitation related to the mirror thickness and minimum spacing \cite{Sve05}.

One possible solutions to
improving angular resolution is to invoke the typical use case of the standard lobster eye configuration as an All Sky Monitor (ASM) for X-ray Astronomy. The lobster eye is used onboard a space-based platform and will continuously scan the sky. If an area of the sky is outside the FoV of the optics, it will be inside the FoV sometime later because of scanning. 
This operational scenario allows for a smaller FoV in the scanning direction, which in-turn permits finer angular resolution.
The desired optics would have a wide FoV and moderate angular resolution in one direction, and a smaller FoV and better angular resolution in another.

It is necessary to use curved mirrors to achieve better angular resolution. However, this puts constraints on the mirror dimensions. A combination of the standard one-dimensional lobster eye optics in one direction and  K-B parabolic mirrors in the other direction meet the desired requirements \cite{Sve05}, shown in Fig. 25.

Preliminary results of this configuration indicate that the Hybrid lobster eye works as intended, i.e. it improves the angular resolution in one direction while still having a wide FoV in another. However, the blurring increases rapidly with the off-axis distance in the direction where there is focusing from the parabolic mirrors. Consequently, it is reasonable to think about such optics for pointed observations if the source and/or image are expected to be highly asymmetric. The effect of blurring is reduced for scanning observations, hence the increase in angular resolution is achievable.
There is a loss of sensitivity with this configuration, which translates to a significant decrease in the limiting flux. This fact, combined with manufacturing difficulties, makes this configuration of limited use for space-based applications. However, there is potential for use in laboratory applications\cite{Sve06b}.

\begin{figure}
\begin{center}
\includegraphics[width=80mm]{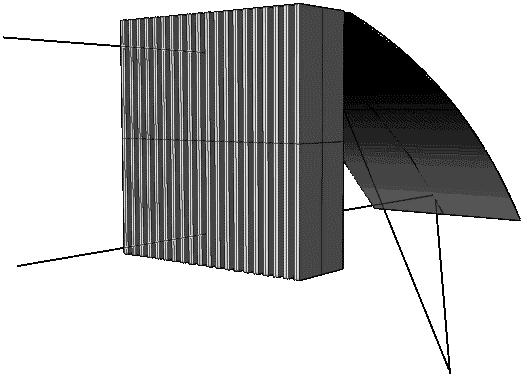}
\caption{The sketch of the Hybrid lobster eye with two plotted rays. Only one parabolic mirror is schematically plotted here. Typically, multiple reflecting surfaces have to be used \cite{Sve05}.
}
\label{Fig2}
\end{center}
\end{figure}

\subsection{Space experiments with lobster eye MFO X-ray optics}

The lobster eye optics in the Schmidt/MFO arrangement was placed on-board the Czech nanosatellite VZLUSAT-1 and on-board the NASA Water Recovery Rocket experiment. More systems are in study and/or in preparation for future space missions. For example, the HORUS double test module was designed and tested recently in order to compare modules with various reflective layers, see Fig. 24 \cite{Stehlik21}. 

\subsubsection{VZLUSAT-1}
The small lobster eye telescope onboard the
VZLUSAT–1 nanosatellite uses the first lobster eye MFO Schmidt X–ray optics in space. 

 The first Czech technological CubeSat satellite VZLUSAT-1 was designed and built during the 2013 to 2016 period. It was successfully launched into Low Earth Orbit at an altitude of 505 km  on June 23, 2017 as part of international mission QB50 onboard a PSLV C38 launch vehicle. The satellite was developed in the Czech Republic by the Czech Aerospace Research Centre, in cooperation with Czech industrial partners and universities \cite{Daniel16}. 
 
 The payload fits into a 2U CubeSat (extended to 3U in space) and includes a 1D \cite{Pina15} \cite{Pina16} miniature X-ray telescope with a Timepix detector in its focal plane \cite{Baca16}.
 
The main mission goal is the technological verification of the system \cite{Urban17}
\cite{Daniel16}. However, there is potential for science as the telescope will view bright celestial sources as part of its mission \cite{Blazek}.
The satellite represents the 5th satellite in space
with Czech X–ray optics onboard.
The 1D lobster eye module onboard VZLUSAT–1 has focal length of 250 mm
and is composed of 116 wedges and 56 reflective double-sided gold-plated foils
(thickness 145 microns). The input aperture is 29×19 mm$^{2}$, outer dimensions are
60×28×31mm$^{3}$. The active part of the foils is 19mm in width and 60 mm in length
and the energy range is 3 to 20 keV. Images of the optics are shown in Fig. 26.

\begin{figure}
\begin{center}
\includegraphics[width=48mm]{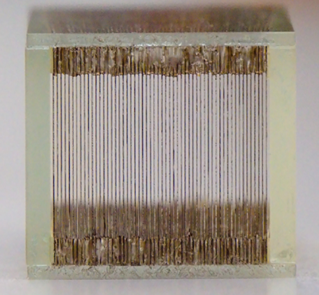}
\includegraphics[width=55mm]{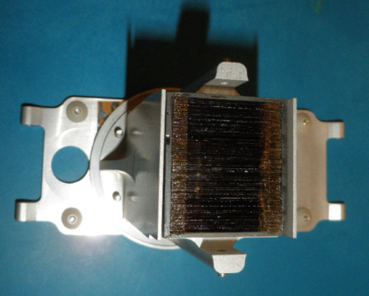}
\caption{The miniature 1D Schmidt lobster eye module for VZLUSAT1 CubeSat \cite{Urban17}.}
\label{Fig2}
\end{center}
\end{figure}

\begin{figure}
\begin{center}
\includegraphics[width=60mm]{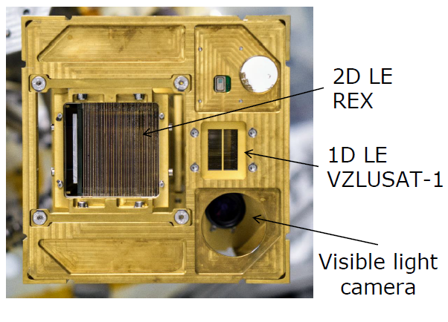}
\includegraphics[width=40mm]{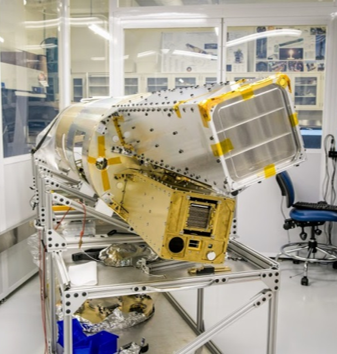}
\caption{The 1D and 2D Schmidt lobster eye modules REX for rocket flight experiment\cite{Urban21}.}
\label{Fig2}
\end{center}
\end{figure}

\subsubsection{REX Rocket Experiment}

The Rocket EXperiment 1 (REX1) was a secondary payload instrument on the Water Recovery X-ray Rocket (WRX-R) experiment. WRX-R was launched from the Kwajalein Atoll in the Marshall Islands on 4
April 2018. WRX-R was the first astrophysics sounding rocket mission to use a newly developed NASA water recovery system for astronomical payloads as an cost effective alternative to typical land recoveries that also may result in payload damage \cite{WRX}.

The WRX-R was led by the Pennsylvania State University (PSU), USA. The primary payload
was the soft X-ray spectroscope of PSU. WRX-R's primary instrument was a grating spectrometer that consisted of
a mechanical collimator, an X-ray reflection grating array, a grazing incidence mirrors, and a hybrid CMOS detector.
The Czech team provided the REX1 optical instrument as a secondary payload \cite{Urban21}
\cite{Daniel17}
\cite{Daniel19}. It was the first time that an X-ray lobster eye telescope was flown in a rocket experiment to observe an astrophysical object. The
design of the REX1 instrument for the WRX-R was based on the concept of an optical baffle, which
is normally used for NASA Sounding rocket experiments. This is a simple construction of a quill-shaped boulder
with the anchor on one side of the block base, where the baffle is attached to the sounding rocket.

The REX1 optical instrument consisted of two parts - vacuum chamber and hermetically sealed box. The vacuum part
contained two (one 1D and one 2D) X-ray telescopes with Timepix pixel detectors \cite{Pina19}. The modules were assembled
using Multi-Foil Technology (MFT). The material of the housing of the optical module was an aluminum alloy.
The 1D X-ray lobster eye system with a focal length of 250 mm, had a FoV of 3.3 x 2.0 degrees and spanned the spectral range from 3 keV to 20 keV. The 1D lobster eye module was composed of 116 wedges and 56
reflective double-sided gold-plated glass foils (thickness of 145 $\mu$m). The gold coating allows the material to
reflect incoming X-ray photons that have shallow incident angles of 0.5 deg or less. The input aperture was 29 x 19
mm$^{2}$, while the outer dimensions were 60 x 28 x 31 mm. The active area of the module was 19 mm in width and
6 mm in length and the energy range was 3 to 20 keV.

The second lobster eye telescope was a 2D X-ray system with a focal length of 1065 mm. The FoV of
this system was 0.8 x 0.8 deg with spectral range from 3 keV to 10 keV. The 2D lobster eye X-ray optics of REX was
composed of two 1D sub-modules where one-sided gold-plated glass foils were in the vertical plane of the horizontal arrangement. Each sub-module consisted of 55 pieces of thin 
at glass foils (thickness of 0.34 mm) which were
arranged so that the focal length was around 1.0 meter. The external dimensions of the module was approximately
80 x 80 x 170 mm. Both REX1 lobster eye modules can be seen in the Fig. 27.

The 2nd generation of the optical system for the Rocket Experiment (REX2) is currently under study \cite{Pina21}. This optical device
is based on the successful mission REX1 described above. The purpose of REX2 is to verify the X-ray optical system that consists of a wide-field 2D X-ray lobster eye assembly with an uncooled Quad Timepix3 detector (512x512 px @ 55 microns and
spectrometer (active area 7 mm$^{2}$, resolution 145 eV @ 5.9 keV ). The 2D X-ray lobster eye optics is a combination
of two 1D lobster eye modules with a focal length of up to 1 m and a FoV larger than 4.0 x 4.0 deg. The
proposed optical system has imaging capabilities (2.5 to 20 keV) and spectroscopy capabilities (0.2 to 10 keV).
The optical system was recently tested in an  X-ray vacuum chamber \cite{Pina21}.



\section{Kirkpatrick-Baez Optics}
\label{KBop}

In this section we briefly describe  Kirkpatrick-Baez (K-B) X-ray optics. From the standpoint of manufacturing, there is significant number of similarities to lobster eye optics in MFO Schmidt arrangements as both are based on multiple thin foils.

Although the Wolter systems are generally well known, Hans Wolter was not the first who proposed X-ray imaging systems based on the reflection of X-rays. In fact, the first grazing incidence system to form a real image was proposed by Kirkpatrick and Baez in 1948\cite{kirk}. This system consists of a set of two orthogonal parabolas in the configuration shown in Figure 28. The first reflection focuses to a line, which the second surface focuses to a point. This was necessary to avoid the extreme astigmatism suffered by a single mirror but was still not free from geometric aberrations. The system is nevertheless attractive for the ease of constructing the reflecting surfaces. These surfaces can be produced as flat plates and then mechanically bent to the required curvature. In order to increase the aperture, a number of mirrors can be nested together, but it should be noted that such nesting introduces additional aberrations.
This configuration is used mostly in experiments not requiring large collecting area (solar, laboratory). Recently, however, large modules of K-B mirrors have been suggested also for stellar X-ray experiments \cite{Hud18} \cite{Hud2018b}.

\begin{figure}
\begin{center}
\includegraphics[width=100mm]{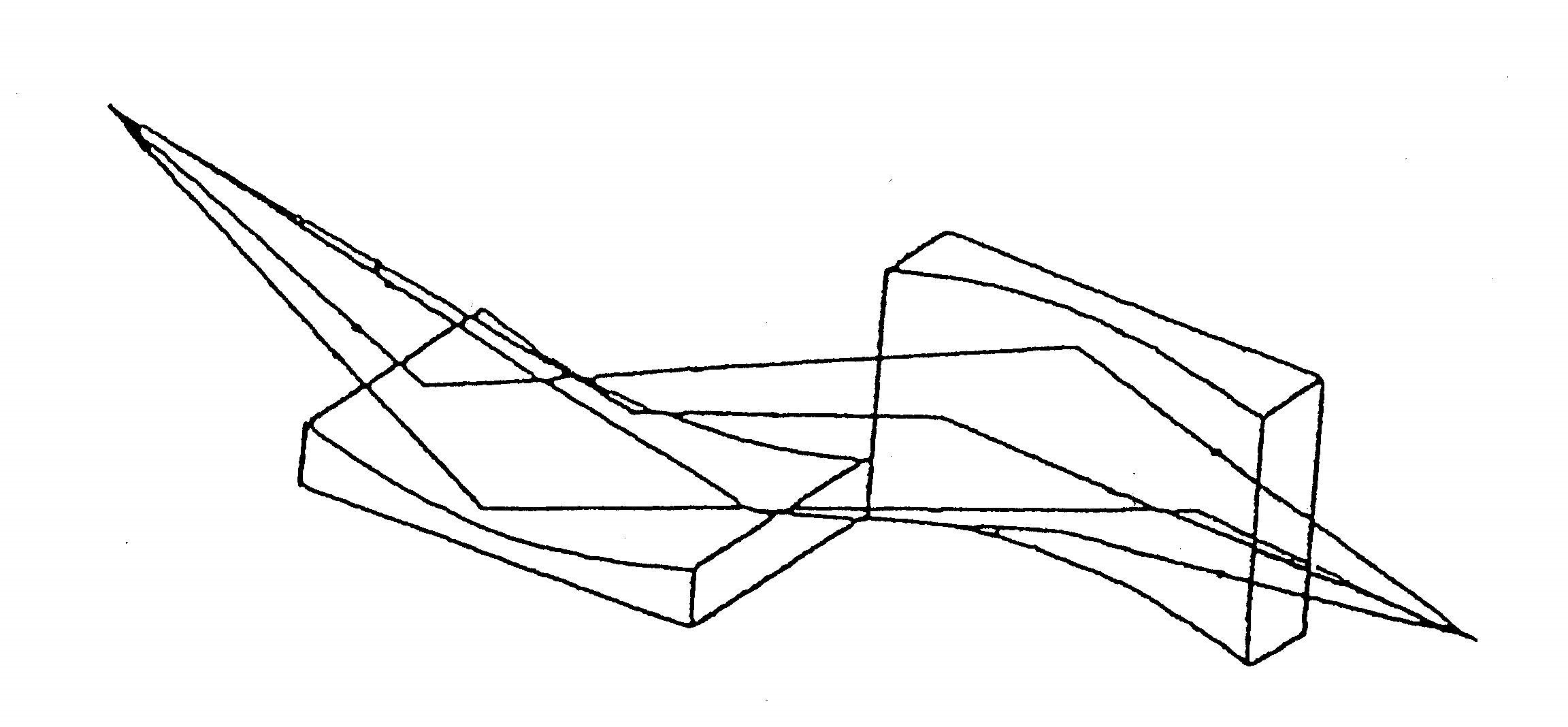}
\caption{The configuration of the K-B X-ray objective according to Kirkpatrick and Baez, 1948 \cite{kirk}).
}\label{Fig2}
\end{center}
\end{figure}

\begin{figure}
\begin{center}
\includegraphics[width=45mm]{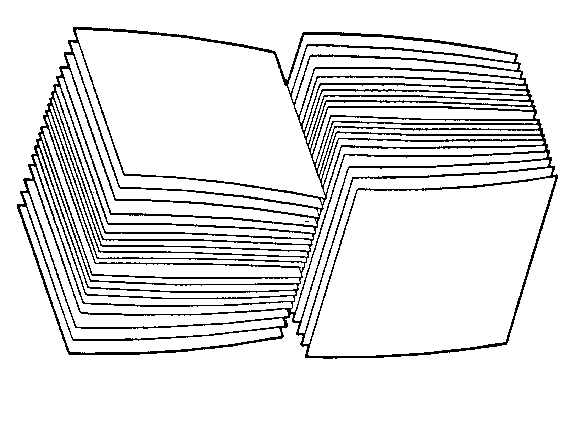}
\includegraphics[width=55mm]{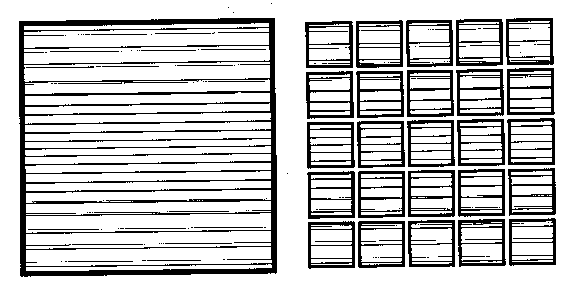}
\caption{Kirkpatrick-Baez mirror consisting of orthogonal stacks of reflectors. Each reflector is a parabola in one dimension. A large K-B mirror can be segmented into rectangular modules of equal size and shape \cite{Goren96}.}
\label{Fig2}
\end{center}
\end{figure}

\begin{figure}
\begin{center}
\includegraphics[width=50mm]{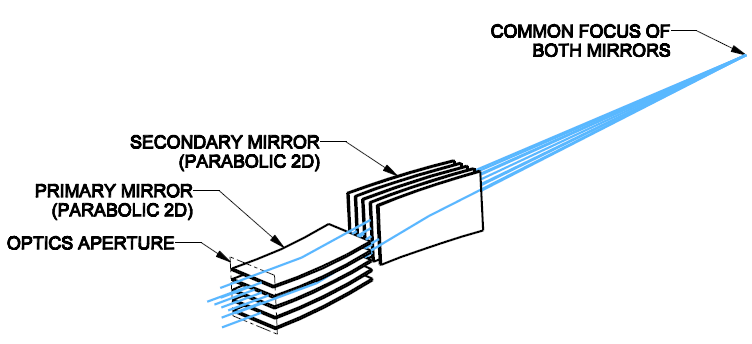}
\includegraphics[width=50mm]{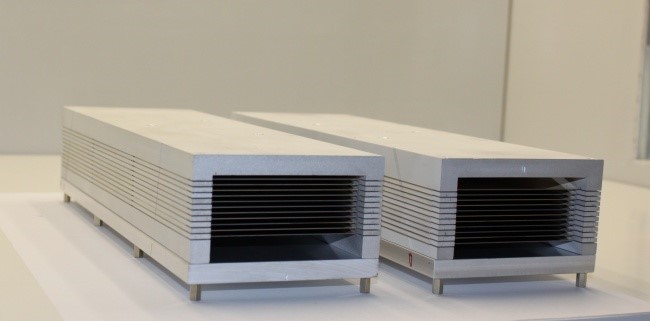}
\caption{Principle of K-B MFO telescope (left \cite{Marsik09})
Laboratory samples of advanced K-B MFO modules designed and developed at Rigaku Innovative Technologies Europe (RITE) in Prague (right \cite{Marsik09}, photo courtesy Rigaku).}
\label{Fig2}
\end{center}
\end{figure}



Despite this fact, astronomical X-ray telescopes flown so far on satellites mostly used the Wolter 1 type optics. However, K-B  was used in several rocket experiments in the past, and in addition to that, they were proposed and discussed for use on several satellite experiments. Alternately, in the lab, K-B systems are in frequent use, e.g. at synchrotron facilities. 

In order to increase the collecting area (the frontal area), a stack of parabolas of translation can be constructed for astrophysical applications. However, in contrast to the single double-plate system, the image of a point-like source starts to become increasingly extended in size as the number of plates involved increases. Wolter type I telescopes bend the incident ray direction two times in the same plane, whereas the two bendings in K-B systems occur in two orthogonal planes, which for the same incidence angle on the primary mirror requires a longer telescope \cite{Aschen}. 


\subsection{K-B systems in astronomical applications}


As an alternative to Wolter optics based instruments, Van Speybroeck et al.\cite{Van} designed several K-B telescope configurations that focus the x rays with sets of two orthogonal parabolas of translation.  According to Van Speybroeck et al.\cite{Van}, the crossed parabola systems should find application in astronomical observations such as high sensitivity surveys, photometry, and certain kinds of spectroscopy where a large effective area rather than high angular resolution is the most important factor.

The design of a K-B grazing incidence X-ray telescope to be used to scan the sky, would allow for the distribution of the reflected X rays and spurious images over the FoV to be analyzed. Kast\cite{Kast} has shown that in order to obtain maximum effective area over the FoV, it is necessary to increase the spacing between plates for a scanning telescope as compared to a pointing telescope. Spurious images are necessarily present in this type of lens, but they can be eliminated from the FoV by adding properly located baffles or collimators.

X-ray telescopes of the type suggested by Kirkpatrick and Baez \cite{kirk} have several advantages over other types of X-ray telescopes for a general sky survey for low-energy X-ray sources. These telescopes use two orthogonal sets of nested parabolas of translation  (perpendicular to one another) to provide 2D focusing of an X-ray image. Although their angular resolution for axial rays is somewhat worse compared with telescopes using successive concentric figures of revolution, they can be constructed more easily and have greater effective area \cite{Van}. Note that more recent papers give somewhat different findings, namely that the K-B Si stacks provide an alternative solution with a reduced
on-axis collecting area but wider field of view and comparable angular resolution \cite{Willi10}.  In either case, these telescopes, in general, can be constructed more easily. The design of K-B-type telescopes has been discussed by several authors e.g. \cite{Van} \cite{Goren73}\cite{Weissk}, and results have been reported from several experiments using 1D focusing from a single set of plates \cite{Goren71}\cite{Catura}\cite{Borken}. For a more recent status see \cite{Hud10} \cite{Hud18}.

\subsubsection{K-B as a segmented mirror}

Segmentation can also be applied to an array of K-B stacked  orthogonal parabolic reflectors (Figure 29). As shown in Figure 29, a large K-B mirror can be segmented into rectangular modules of equal size and shape \cite{Goren96}. A segmented K-B telescope has the advantage of being highly modular on several levels. All segments are rectangular boxes with the same outer dimensions. Along a column, the segments are nearly identical and many are interchangeable with each other. All reflectors deviate from flatness only slightly. On the other hand, the Wolter reflectors are highly curved in the azimuthal direction and the curvature varies over a wide range. Furthermore, within a segment, the K-B reflectors themselves can be segmented along the direction of the optical axis. As shown in Figure 29, a K-B mirror system can be folded more easily than the Wolter mirror into a compact volume for launch and deployment in space. The examples of assembled K-B modules based on superior quality  gold coated Si wafer substrates are illustrated in Figure 30.

\subsubsection{K-B in Astronomical Telescopes: Recent Status and Future Plans}

The first attempt to create an astronomical K-B module with silicon wafers was reported by Joy et al. \cite{Joy}. A telescope module that consisted of 94 silicon wafers with diameter of 150 mm, uncoated, with thickness of  0.72 mm was constructed. The device was tested both with optical light and with X-rays. The measured FWHM was 150 arc-seconds, which was dominated by large-scale flatness. It should be noted that the surface quality and flatness of Si wafers has improved since this time.

Recent efforts towards supporting future larger and precise imaging astronomical X-ray telescopes require re-considering both the technologies and mirror assembly design. Future large X-ray telescopes require new light-weight and thin materials/substrates such as glass foils and/or silicon wafers\cite{Hud15c}. 
Their shaping to small radii, as required in Wolter designs, is not an easy task. While the K-B arrangements have potential to represent a less laborious and hence less expensive alternative because of (i) no need of mandrels (ii) no need of polishing and (iii) no need of bending to small radii.

The use of K-B arrangement for the proposed IXO project (the proposed joint NASA/ESA/JAXA International X-ray Observatory) was suggested and investigated by Marsikova et al.\cite{Marsik09}, Hudec et al. \cite{Hud2011}, and by Willingale and Spaan \cite{Willi10}. These investigations indicate that if superior quality reflecting plates were used and the focal length is large, an angular resolution of order of a few arcsec could be achieved. Recent simulations further indicate that in comparison to Wolter arrangement, the K-B optics exhibit reduced on axis collecting area but larger FoV, at comparable angular resolution \cite{Willi10}. 


A very important factor is the ease of constructing highly segmented modules based on multiply nested thin reflecting substrates if compared with Wolter design. While e.g. the Wolter design for future large space X--ray telescopes such as Athena requires the substrates to be precisely formed with curvatures as small as 0.25 m, the alternative K-B arrangement uses almost flat or only slightly bent sheets.

Hence, the feasibility to construct a K-B module with the required Athena 5 arc-second HEW resolution at an affordable cost is, in principle, lower than the cost of a Wolter arrangement.
Note however that in order to achieve the comparable effective area, the focal length of K-B system is required to be about twice of the focal length of Wolter system\cite{Marsik09}\cite{Hud10}\cite{Hud2011}.


\section{Summary}
\label{sum}

The grazing incidence X-ray optical elements of non-Wolter type (lobster eye and Kirkpatrick-Baez) offer alternative solutions for many future space- and lab-based applications. They can offer cheaper, and/or lighter alternatives, and also a much larger FoV. At the same time, new computer-based systems allow us to consider alternative designs and arrangements \cite{Nentv17}.

Although both Angel and Schmidt designs were suggested in the 70's, both have seen rapid development over the past few years with MPO optics in an Angel arrangement already on selected missions and the Schmidt design using MFOs being proven on rocket and CubeSat experiments.

A direct and reliable comparison between MFO and MPO designs of lobster eye X-ray optics is difficult, as
in both cases the real optics performance deviates from the theoretical. The necessary slumping of the MPOs introduces additional sources of error\cite{Bannister,dickspie}, whilst the MFO design is harder to assemble. Both designs differ in geometry using both Angel and Schmidt designs, and require different manufacturing and assembling technology. The MFO technology enables a larger effective area with easy deposition of reflective layers, whilst the MPOs are lighter and are easier to assemble into a large array. The effective area at 10 keV for MFOs is higher than for MPOs although alternative coatings are being investigated for MPOs to improve the higher energy response.

The prototypes developed and tested for both arrangements confirm that these light weight telescopes are fully feasible and can achieve angular resolutions of several arcmin or better over a very wide FoV. While both provide a more modest angular resolution compared to Chandra\cite{1} and XMM-Newton\cite{2} for example, they can still be used to help solve pressing questions in X-ray astrophysics, and can also be used for other applications such as within laboratories. K-B optics have already found wide applications in synchrotrons, and have demonstrated their performance and advantages.



\section{Acknowledgements}
The authors wish to thank the other members of their research groups. The research leading to these results has received funding from the European Union’s Horizon 2020 Programme under the AHEAD2020 project (grant agreement n. 871158)

\bibliography{bobs, bobs2}
\bibliographystyle{spbasic}
\end{document}